%% file: TOP-14-012_temp.tex
\pdfoutput=1

\documentclass[11pt,twoside,a4paper,cmspaper,final,collab]{cms-tdr}
\def\svnVersion{370771}\def\cmsCernNoTag{CERN-EP/2016-078}\def\cmsCernDate{\today}\def\cmsMessage{Published in Physical Review D as \href{http://dx.doi.org/10.1103/PhysRevD.94.072002}{\doi{10.1103/PhysRevD.94.072002.}}}

\begin{document}\cmsNoteHeader{TOP-14-012}

\hyphenation{had-ron-i-za-tion}
\hyphenation{cal-or-i-me-ter}
\hyphenation{de-vices}
\RCS$HeadURL: svn+ssh://svn.cern.ch/reps/tdr2/papers/TOP-14-012/trunk/TOP-14-012.tex $
\RCS$Id: TOP-14-012.tex 370676 2016-10-14 09:49:32Z skinnari $

\newlength\cmsFigWidth
\ifthenelse{\boolean{cms@external}}{\setlength\cmsFigWidth{0.85\columnwidth}}{\setlength\cmsFigWidth{0.4\textwidth}}
\ifthenelse{\boolean{cms@external}}{\providecommand{\cmsLeft}{top}}{\providecommand{\cmsLeft}{left}}
\ifthenelse{\boolean{cms@external}}{\providecommand{\cmsRight}{bottom}}{\providecommand{\cmsRight}{right}}
\ifthenelse{\boolean{cms@external}}{\newcommand{\cmsTable}[2]{\relax#2}}{\newcommand{\cmsTable}[2]{\resizebox{#1}{!}{#2}}}

\providecommand{\Pj}{\ensuremath{\cmsSymbolFace{j}}\xspace}
\providecommand{\PPsix}{\POWHEG{}+\PYTHIA{}6\xspace}
\providecommand{\MCATHsix}{\MCATNLO{}+\HERWIG{}6\xspace}
\providecommand{\MADPsix}{\MADGRAPH{}+\PYTHIA{}6\xspace}
\newcommand{\x}{\ensuremath{\phantom{0}}}
\newcommand{\xx}{\ensuremath{\phantom{00}}}
\newcolumntype{x}{D{,}{\,\pm\,}{3}}

\cmsNoteHeader{TOP-14-012}
\title{Measurement of the integrated and differential \texorpdfstring{\ttbar}{t-tbar} production cross sections for \texorpdfstring{high-$\pt$}{high-pt} top quarks in \texorpdfstring{$\Pp\Pp$}{pp} collisions at \texorpdfstring{$\sqrt{s}=8\TeV$}{sqrt(s) = 8 TeV}}

\date{\today}

\abstract{
The cross section for pair production of top quarks ($\ttbar$) with high transverse momenta is measured in $\Pp \Pp$ collisions, collected with the CMS detector at the LHC with $\sqrt{s} = 8\TeV$ in data corresponding to an integrated luminosity of 19.7\fbinv. The measurement is performed using lepton+jets events, where one top quark decays semileptonically, while the second top quark decays to a hadronic final state. The hadronic decay is reconstructed as a single, large-radius jet, and identified as a top quark candidate using jet substructure techniques. The integrated cross section and the differential cross sections as a function of top quark $\pt$ and rapidity are measured at particle level within a fiducial region related to the detector-level requirements and at parton level. The particle-level integrated cross section is found to be $\sigma_{\ttbar} = 0.499 \pm 0.035\,(\text{stat+syst}) \pm 0.095\thy \pm 0.013\lum\unit{pb}$ for top quark $\pt > 400\GeV$. The parton-level measurement is $\sigma_{\ttbar} = 1.44 \pm 0.10\,(\text{stat+syst}) \pm 0.29\thy\pm 0.04\lum\unit{pb}$. The integrated and differential cross section results are compared to predictions from several event generators.
}

\hypersetup{
pdfauthor={CMS Collaboration},
pdftitle={Measurement of the integrated and differential t-tbar production cross section for high-pt top quarks in pp collisions at sqrt(s)=8 TeV},
pdfsubject={CMS},
pdfkeywords={CMS, physics, top physics}}

\maketitle

\section{Introduction}
\label{sec:introduction}

Measurements of top quark pair (\ttbar) production cross sections provide crucial information for testing the standard model (SM) and the accuracy of predictions from Monte Carlo (MC) generators. The CMS~\cite{CMS_JINST} and ATLAS~\cite{ATLAS_JINST} Collaborations at the CERN LHC have previously measured the differential $\ttbar$ cross sections at $\sqrt{s} = 7$ and 8\TeV as a function of transverse momentum (\pt) and other kinematic properties of the top quarks and the overall $\ttbar$ events~\cite{diff_xs_ttbar_resolved, diff_xs_ttbar_resolved_8TeV, Khachatryan:2015fwh, Aad:2012hg, Aad:2014zka, Aad:2015eia, Aad:2015mbv}. These measurements use events where each parton from the top quark decay is associated with a distinct jet. However, when top quarks are produced with large Lorentz boosts, their decays are often collimated and the final decay products may be merged.
For a top quark with a Lorentz boost of $\gamma = E/m$, where $E$ is the energy and $m$ the mass of the top quark, the angle $\Delta R$ in radians between the $\PW$ boson and the \PQb quark from the top quark decay is approximately $\Delta R = 2/\gamma$.
In this paper, a measurement of the $\ttbar$ production cross section is presented utilizing jet substructure techniques to enhance sensitivity in the kinematic region with high-$\pt$ top quarks.
Accurate modeling of the boosted top quark regime is important as it is sensitive to many physics processes beyond the SM, as discussed, for example, in Ref.~\cite{Frederix:2007gi}.

This paper presents the first CMS measurement of the $\ttbar$ production cross section in the boosted regime. The cross section is measured as a function of the top quark transverse momentum ($\pt^{\PQt}$) and rapidity ($y^{\PQt}$) for $\pt^{\PQt} > 400\GeV$, corresponding to the upper $\pt$ range covered by the CMS measurement in Ref.~\cite{diff_xs_ttbar_resolved_8TeV}. A dedicated measurement of $\ttbar$ production in the boosted regime has recently  been reported by the ATLAS Collaboration~\cite{Aad:2015hna}.

The analysis is performed for events in lepton+jets final states where one top quark decays according to
$\PQt \to \PW \PQb \to \ell \PGn \PQb$, with $\ell$ denoting an electron or a muon, and the second top quark decays to quarks ($\PQt \to \PW \PQb \to \PQq \PAQq' \PQb$).
Lepton+jets final states originating from $\PW$ boson decays to $\tau$ leptons ($\PQt \to \PW \PQb \to \tau \PGn \PQb \to \ell \PAGn \PGn \PQb$) are treated as background.
The boosted top quark that decays to a hadronic final state is reconstructed as a single, large-radius (large-$R$) jet.
Jet substructure techniques similar to those used in Refs.~\cite{B2G-13-001,B2G-13-008} are applied to identify those large-$R$ jets originating from top quarks ($\PQt$-tagged jets).
A maximum-likelihood fit is performed to extract the background normalizations, the $\PQt$ tagging efficiency, and the integrated $\ttbar$ production cross section for $\pt^{\PQt} > 400\GeV$.
The results are presented at the particle level in a fiducial region similar to the event selection criteria to minimize the dependence on theoretical input, and fully corrected to the parton level. Differential \ttbar cross sections are also measured at the particle (parton) level as a function of the $\PQt$-tagged jet (top quark) \pt and $y$ after subtracting the background contributions and correcting for inefficiencies and bin migrations.

\section{The CMS detector, event reconstruction, and event samples}
\label{sec:samples}

The CMS detector~\cite{CMS_JINST} is a general-purpose detector that uses a silicon tracker, a finely segmented lead tungstate crystal electromagnetic calorimeter (ECAL), and a brass and scintillator hadron calorimeter (HCAL). These subdetectors have full azimuthal coverage and are contained within the bore of a superconducting solenoid that provides a 3.8\unit{T} axial magnetic field.  Char\-ged particles are reconstructed in the tracker, covering a pseudorapidity~\cite{CMS_JINST} range of $\abs{\eta} < 2.5$. The surrounding ECAL and HCAL provide coverage for photon, electron, and jet reconstruction for $\abs{\eta} < 3$.
Muons are measured in gas-ionization detectors embedded in the steel flux-return yoke outside the solenoid. Events are reconstructed using the particle-flow algorithm~\cite{CMS-PAS-PFT-09-001,CMS-PAS-PFT-10-001}, which identifies each particle with an optimized combination of all subdetector information.
The missing transverse momentum vector \ptvecmiss is defined as the projection on the plane perpendicular to the beams of the negative vector sum of the momenta of all reconstructed particles in an event. Its magnitude is referred to as \ETmiss.
A more detailed description of the CMS detector, together with a definition of the coordinate system used and the relevant kinematic variables, can be found in Ref.~\cite{CMS_JINST}.

The measurement is performed using the CMS data recorded at $\sqrt{s} = 8\TeV$, corresponding to an integrated luminosity of $19.7 \pm 0.5\fbinv$~\cite{LumiUncertainty}. For the $\Pe$+jets channel, data are collected with a trigger requiring an electron with $\pt > 30\GeV$ and $\abs{\eta}<2.5$, at least one jet with $\pt > 100\GeV$, and at least one additional jet with $\pt > 25\GeV$. For the $\mu$+jets channel, the trigger demands a muon with $\pt > 40\GeV$ and $\abs{\eta} < 2.1$, with no jet requirements. At the trigger level, the leptons are not required to be isolated.

Simulated events are used to estimate the efficiency to reconstruct the $\ttbar$ signal, evaluate the systematic uncertainties, and model most of the background contributions. Samples of $\ttbar$ and electroweak single top quark events are generated using the next-to-leading-order (NLO) MC generator \POWHEG (v. 1.0)~\cite{powheg01, powheg02, powheg, powheg_ttbar, powheg_singletop}, while $\PW$ boson production in association with jets is generated with the leading-order (LO) generator \MADGRAPH (v. 5.1.3.30)~\cite{MadGraph}. Additional $\ttbar$ samples, generated using {\MADGRAPH} and the NLO generator {\MCATNLO} (v. 3.41)~\cite{mcnlo}, are used for comparison with {\POWHEG}.
The {\MCATNLO} production is interfaced to {\HERWIG} (v. 6.520, referred to as {\HERWIG{}6} in the following)~\cite{herwig} for parton showering, while all other generators are interfaced to \PYTHIA (v. 6.426, referred to as \PYTHIA{}6)~\cite{pythia}.
For the samples produced with {\MADGRAPH}, the MLM prescription~\cite{Mangano:2006rw} is applied for matching of matrix-element jets to parton showers.
The most recent \PYTHIA Z2* tune is used. It is derived from the Z1 tune~\cite{Field:2010bc}, which uses the CTEQ5L parton distribution function (PDF) set, whereas Z2* adopts CTEQ6L~\cite{cteq}.
The \POWHEG $\ttbar$ and single top quark samples are generated using the CT10 next-to-next-to-leading-order (NNLO)~\cite{Gao:2013xoa} PDFs, while the {\MCATNLO} $\ttbar$ sample uses the NLO CTEQ6M~\cite{cteq} PDF set.
The LO CTEQ6L1~\cite{cteq} PDF set is used for the \MADGRAPH $\ttbar$ and $\PW$+jets samples.
All generated events are propagated through a simulation of the CMS detector based on \GEANTfour (v. 9.4)~\cite{Geant4}.

The simulated events are corrected to match the conditions observed in data. All simulated events are reweighted to reproduce the distribution of the number of primary vertices that arises from additional $\Pp\Pp$ interactions within the same or neighboring bunch crossings (pileup), as measured in data. The jet energy resolution is corrected by scaling the difference between the generated and the reconstructed jet momentum so that the resolution matches that observed in data~\cite{jme_jinst}. Lepton trigger and identification efficiencies are also corrected for differences between data and simulation. Jet energy corrections are obtained from the simulation and further corrections are applied to data from in situ measurements using the energy balance in dijet and photon+jet events~\cite{jme_jinst}. The contribution to the jet energy in data from pileup is removed using the area-based subtraction technique outlined in Ref.~\cite{jetarea_fastjet}, augmented by corrections from data as a function of the jet $\eta$, as described in Ref.~\cite{jme_jinst}.

\section{Event selection}
\label{sec:selection}

Jet clustering is performed with the {\FASTJET} package (v. 3.1)~\cite{fastjet}. Two jet clustering algorithms are used in the measurement. The anti-\kt algorithm~\cite{antikt} with a distance parameter $R=0.5$ is used to reconstruct jets that are hereafter referred to as small-$R$ jets.
Lepton candidates that are found within $\Delta R < 0.5$ of a jet, where $\Delta R = \sqrt{\smash[b]{{(\Delta \eta)}^2 + {(\Delta \phi)}^2}}$ and $\Delta \eta$ and $\Delta \phi$ are the pseudorapidity and azimuthal angle (in radians) differences between the direction of the lepton and the jet, are subtracted from the jet four-vector to avoid including such leptons within jets.
The small-$R$ jets are required to have $\pt>30\GeV$ and $\abs{\eta} < 2.4$.
Small-$R$ jets that are identified as originating from a bottom (b) quark through the use of an algorithm that combines secondary-vertex and track-based lifetime information~\cite{btag_paper,BTV-13-001} are classified as being $\PQb$ tagged.
The algorithm working point used has an efficiency for tagging a $\PQb$ jet of ${\approx}65\%$, while the probability to misidentify light-flavor jets as $\PQb$ jets is ${\approx}1.5\%$.
The secondary-vertex mass of the $\PQb$-tagged jet ($m_\text{vtx}$) is defined as the invariant mass of the tracks associated with the secondary vertex, assuming that each particle has the pion mass. Jets that are $\PQb$ tagged are also required to have a secondary vertex (resulting in a small change in the efficiency). Differences in $\PQb$ tagging efficiency and misidentification rates between data and simulated events are accounted for through scale factors applied to the simulation.

The second jet clustering algorithm is the Cambridge--Aachen (CA) algorithm~\cite{CAaachen,CAcambridge}, used to reconstruct large-$R$ jets with a distance parameter $R=0.8$. These jets are required to have $\pt > 400\GeV$, where this lower $\pt$ bound is set such that the top quark decay products are typically fully merged for $R=0.8$. The kinematics of the large-$R$ jet is used for the $\pt^{\PQt}$ and $y^{\PQt}$ measurements.

The CMS top quark tagging algorithm~\cite{JME-13-007}, using large-$R$ jets as input, is employed in this measurement to identify top quark candidates decaying hadronically. The algorithm begins by identifying subjets through recursive declustering of the original large-$R$ jet, reversing the clustering sequence of the CA algorithm. First, the last clustering step is reversed, splitting the large-$R$ jet $\Pj$, with transverse momentum denoted as $\pt^{\Pj}$, into two subjets $\Pj_1$ and $\Pj_2$, with transverse momenta $\pt^{\Pj_1}$ and $\pt^{\Pj_2}$. If the two subjets satisfy $\Delta R (\Pj_1, \Pj_2) >0.4-0.0004 \, \pt^{\Pj}$, with $\pt^{\Pj}$ in GeV, they are passed to the next step of the algorithm; if not, they are reclustered and the parent is labeled as a hard subjet. Each subjet is required to satisfy $\pt^{j_{i}} > 0.05 \, \pt^{\Pj}$; otherwise, the subjet is discarded. A secondary decomposition is next applied to the subjet(s), identifying up to a maximum of four hard subjets.

The large-$R$ jet that is identified as a $\PQt$ jet candidate is required to contain at least three subjets, corresponding to the presumed $\PQb$, $\PQq$, and $\PAQq'$ fragmentation products. In addition, the minimum pairwise invariant mass of the three subjets of highest $\pt$ is required to be greater than 50\GeV, as expected for the $\PQt \to \PW \PQb$ decay, and the total jet invariant mass $m_{\Pj}$ is required to be consistent with the top quark mass by demanding $140<m_{\Pj}<250\GeV$. Large-$R$ jets which fulfill these requirements are labeled as $\PQt$-tagged jets. The cumulative efficiency for these $\PQt$ tagging requirements is about 25\% for $\abs{\eta}<1.0$ and 13\% for $1.0<\abs{\eta}<2.4$~\cite{JME-13-007}. The difference in the $\PQt$ tagging efficiency between data and simulation is accounted for through a scale factor applied to the simulation that is derived using a maximum-likelihood fit.

Electrons~\cite{Khachatryan:2015hwa} and muons~\cite{Chatrchyan:2012xi} must have, respectively, $\pt > 35\GeV$ and 45\GeV, and $\abs{\eta} < 2.5$ and 2.1, where the differences are a consequence of the requirements on the respective lepton triggers. Since leptons from high-$\pt$ top quark decays are often emitted close to their accompanying $\PQb$ jets, they may not be well-isolated. To reject background contributions from jets misidentified as leptons, the leptons must pass a two-dimensional (2D) selection, requiring either $\Delta R (\ell,$ closest small-$R$ jet$) > 0.5$ or $\pt^\text{rel} > 25\GeV$, where $\pt^\text{rel}$ is the component of the lepton $\pt$ perpendicular to the axis of the closest small-$R$ jet.
An additional criterion is applied in the electron channel to further reduce the multijet background contribution from mismeasured jets. The requirement ensures that $\ptvecmiss$ does not point parallel to the direction of either the electron ($\Pe$) or the highest-$\pt$ jet ($\Pj$) for low-$\ETmiss$ events: $\abs{\Delta\phi(\{\Pe~\text{or}~\Pj\},\ptvecmiss)-1.5} < \ETmiss/50\GeV$. Events that contain more than one lepton with $\pt > 20\GeV$ and $\abs{\eta} < 2.5$ (2.1) for electrons (muons) are rejected.

Events selected for the analysis must contain exactly one electron or muon, at least one small-$R$ jet near the lepton ($\Delta R(\ell, \text{jet}) < \pi/2$, referred to as the leptonic side), and one large-$R$ jet away from the lepton ($\Delta R(\ell, \text{jet}) > \pi/2$, referred to as the hadronic side).
These events are next separated into three exclusive event categories with different signal and background admixtures: ``0t", ``1t+0b", and ``1t+1b".
The 0t events are defined by requiring that no hadronic-side jet pass the $\PQt$ tagging selection. For the 1t+0b events, the hadronic-side jet must pass the $\PQt$ tagging selection, and no leptonic-side jets can be $\PQb$ tagged. The third category of 1t+1b events must contain both a hadronic-side $\PQt$-tagged jet and a leptonic-side $\PQb$-tagged jet. The 0t sample is dominated by background events, primarily from $\PW$+jets production, while the signal and background yields for the 1t+0b sample are expected to be of comparable size. The 1t+1b sample is dominated by signal events.

\section{Background estimation}
\label{sec:backgrounds}

The dominant sources of background are single top quark production (primarily from the $\PW \PQt$ channel), $\PW$+jets production, and multijet production. In addition, $\ttbar$ events with decays to $\tau$+jets (resulting in either hadronic or leptonic final states) or any other than $\Pe$/$\mu$+jets final states are treated as background in the measurement, and hereafter referred to as "$\ttbar$ other". Other sources of background, including diboson, $\cPZ$+jets, $\PW\PH$, and $\ttbar \PW /\cPZ$ production, were found to be negligible. All background normalizations are extracted through a maximum-likelihood fit discussed in Section~\ref{sec:xsec}, while the signal and all background distributions are modeled using simulation, except multijet production, which is obtained from data. The $\ttbar$ other contribution is constrained to have the same relative normalization as the $\ttbar$ signal in the likelihood fit.

The background from multijet production is estimated using control samples in data. Multijet templates for each event category (0t, 1t+0b, 1t+1b) are extracted using control samples, defined by inverting the 2D lepton-jet separation requirement and subtracting residual contributions (corresponding to 3--15\% of events in the control samples) from $\ttbar$, single top quark, and $\PW$+jets events. An initial multijet background normalization is obtained for each event category from a fit of multijet and other signal and background templates to the $\ETm$ distribution in data.

\section{Systematic uncertainties}
\label{sec:systematics}

Systematic uncertainties in the measurement arise from reconstruction and detector resolution effects, background estimation, and theoretical uncertainty in the modeling of signal. The dominant experimental uncertainty is the uncertainty in the $\PQt$~tagging efficiency. The different sources of systematic uncertainty are described in detail below.

The uncertainty in the $\PQt$ tagging efficiency and the corresponding data-to-simulation correction factor are evaluated in Ref.~\cite{JME-13-007}. Since there is a large overlap between those events and events in the signal region in this measurement, and since the $\PQt$ tagging efficiency is strongly anticorrelated with the $\ttbar$ cross section measurement, the $\PQt$ tagging efficiency and its uncertainty are determined simultaneously with the cross section (see Section~\ref{sec:xsec_fit}). The resulting efficiency is in agreement with the previous measurement~\cite{JME-13-007}.

The uncertainties in jet energy scale are estimated by changing the jet energy as a function of jet $\pt$ and $\eta$ by $\pm1$ standard deviation~\cite{jme_jinst}. These uncertainties, which include differences in jet response between light- and heavy-flavor jets, have been measured for anti-$\kt$ jets with distance parameters of $R=0.5$ and 0.7, but not for $R=0.8$ CA jets. The response of the $R=0.8$ CA jets is estimated in simulation to be within 1\% of the response of $R=0.7$ anti-$\kt$ jets. This is checked by comparing the reconstructed $\PW$ boson mass in data and simulation in moderately boosted $\ttbar$ events (outside of the signal region). An additional 1\% uncertainty is used to account for the small differences observed in these studies.
The jet energy scale uncertainties for $R=0.5$ and $R=0.8$ jets are treated as fully correlated. 

The jet energy resolution is known to be about 10\% worse in data than in simulation, and the resolution is therefore adjusted in simulation, using smearing factors in bins of jet $\eta$~\cite{jme_jinst}. An associated systematic uncertainty is obtained by rescaling the resolution smearing in simulation by $\pm1$ standard deviation. This corresponds to changes in the smearing of about $\pm$(2.4--5.0)\%, depending on $\eta$. The effect of jet mass scale and jet mass resolution were found to be very small compared to those from the jet energy. These are accounted for with the data-to-simulation correction factor.

The uncertainties associated with the jet energy scale and resolution are propagated to the estimation of the \ETmiss. The uncertainty in the modeling of the large-$R$ jet mass, which was measured in Ref.~\cite{SMP-12-019}, is also accounted for through propagating the jet energy uncertainties to the full jet four-vector.

In addition to uncertainties in the distributions, we also consider several normalization uncertainties affecting the signal yield. The uncertainties in background yields are taken into account in the combined signal-and-background maxi\-mum-likelihood fit by changing the $\PW$+jets, single top quark, and multijet normalizations, assuming conservative log-normal prior uncertainties of $\pm 50$\%, $\pm 50$\%, and $\pm 100$\%, respectively. The background normalizations are constrained in the maxi\-mum-likelihood fit, and corresponding background uncertainties extracted as the $\pm1$ standard deviation uncertainties in the fit. In addition, the statistical uncertainty resulting from the finite sizes of the simulated samples are included.
The uncertainty in the measurement of the integrated luminosity of $\pm 2.6$\%~\cite{LumiUncertainty} is also included.

The uncertainty in the pileup modeling is evaluated by varying the total inelastic $\Pp\Pp$ cross section used in the simulation within its uncertainty of $\pm$5\%~\cite{pumodel}. The resulting uncertainty in the cross section measurements is less than 1\%.

Systematic uncertainties from the lepton trigger and corrections to the lepton identification efficiencies that are applied to all simulated events contribute negligibly to the uncertainty in the cross section measurement.
This includes the lepton $\eta$ dependence of these uncertainties.
The uncertainty in the $\PQb$ tagging efficiency~\cite{btag_paper,BTV-13-001} is also considered, but has a negligible impact on the final result since the measurements are performed by combining events in the 1t+0b and 1t+1b event categories.
Uncertainties pertaining to the modeling of the secondary-vertex mass, which is one of the variables used in the maxi\-mum-likelihood fit, are negligible compared to the statistical uncertainty in the sample. 

Theoretical uncertainties in the modeling of the $\ttbar$ events originate from the choice of PDF and renormalization and factorization ($\mu_R$ and $\mu_F$) scales, whose nominal values are chosen to be equal to the momentum transfer $Q$ in the hard scattering, given by $Q^2 = m_\text{top}^2$, where the summation runs over all final-state partons in the event. The uncertainty in the modeling of the hard-scattering process is evaluated using samples where the renormalization and factorization scales are simultaneously changed up ($2Q$) or down ($Q/2$). The uncertainty from the PDF is evaluated using the up and down eigenvector outputs from the NNLO PDF sets CT10~\cite{Gao:2013xoa}, MSTW 2008~\cite{mstw}, and NNPDF2.3~\cite{nnpdf}, following the PDF4LHC prescription~\cite{Alekhin:2011sk,Botje:2011sn}.
An additional theoretical uncertainty is assigned to account for the choice of event generator and parton shower algorithm in extracting the integrated and differential cross sections, evaluated using {\MCATHsix} (see Sections~\ref{sec:xsec_fit} and~\ref{sec:xsec_diff}).

\section{Cross section measurements}
\label{sec:xsec}

The $\ttbar$ signal yield, background normalizations, and $\PQt$ tagging efficiency are extracted simultaneously using a binned, extended maximum-likelihood fit to different templates of several kinematic variables described below. First, the fit is used to determine the integrated $\ttbar$ cross section for $\pt^{\PQt} > 400\GeV$, providing a simultaneous measurement of the cross section with nuisance parameters and constraints on the background yields in the data. The results are then used to obtain the differential $\ttbar$ cross section as a function of $\pt^{\PQt}$ and $y^{\PQt}$. The cross sections are presented at both the particle and parton levels.

\subsection{Maximum-likelihood fit}
\label{sec:xsec_fit}

Three exclusive event categories are used in the maximum-likelihood fit (0t, 1t+0b, 1t+1b), as defined in Section~\ref{sec:selection}. The lepton $\abs{\eta}$ is used as the discriminant for events in the 0t and 1t+0b categories, while $m_\text{vtx}$ is used to discriminate $\ttbar$ events ($\ttbar$ signal and $\ttbar$ other are constrained to the same relative normalization in the fit) from non-$\ttbar$ background in the 1t+1b event category. The electron and muon channels are fitted separately, yielding a total of six categories. The maximum-likelihood fit is performed within the \textsc{theta} framework~\cite{theta}.

Background normalizations and experimental systematic uncertainties are treated as nuisance parameters in the fit, three of which are built into the model as uncertainties in the input distributions, these being the jet energy scale, jet energy resolution, and $\PQt$ tagging efficiency. The event categories for the fit are designed such that the $\PQt$ tagging efficiency is constrained by the relative populations of events in the different categories. The $\ttbar$ cross section and the background normalizations are therefore correlated with these variables. The strongest correlation with the $\ttbar$ cross section is the $\PQt$ tagging efficiency.
A log-normal prior constraint is used for each nuisance parameter that corresponds to a normalization uncertainty, while uncertainties based on the form of the distributions are modeled with a Gaussian prior for the nuisance parameter, which is used to interpolate between the nominal and shifted templates.
The $\Pe$+jets and $\mu$+jets events use common nuisance parameters for all systematic uncertainties and background normalizations, except for multijet backgrounds, which are taken as independent of each other.
The total fitted uncertainties in the background yields are 46\% for single top quark, 7.5\% for $\ttbar$ other, 6.8\% for $\PW$+jets production, and 47\% and 17\%, respectively, for the muon and electron multijet backgrounds.

A correction factor to account for small differences in the $\PQt$ tagging efficiency between data and simulation is also determined through the maximum-likelihood fit. While the dependence of this efficiency correction on the $\PQt$ jet $\eta$ is taken from Ref.~\cite{JME-13-007}, an additional uncertainty to account for a potential dependence of  $\pt^{\PQt}$ is evaluated by performing separate fits for events with $\pt^{\PQt}<600\GeV$ and $>$600\GeV. All other nuisance parameters are required to be the same in both $\pt^{\PQt}$ regions for this check. An additional uncertainty of 17\% is assigned for $\pt^{\PQt}>600$\GeV to account for the $\pt$ dependence, resulting in a total uncertainty in the $\PQt$ tagging efficiency of $\pm$5\% ($\pm$18\%) for $\pt^{\PQt}<600$ ($>$600)\GeV.

The measured normalizations in the signal and background yields, as determined from the maximum-likelihood fit, are given, together with the number of observed events in data, in Table~\ref{tab:counts_postfit}. The electron and muon channels are shown separately. The quoted uncertainties are from the total fit, and include the statistical components, but not the theoretical uncertainties in the $\ttbar$ signal. The total signal and background yields are consistent with the observed number of events in the data within about one standard deviation.

\begin{table*}[htb]
\topcaption{\label{tab:counts_postfit} Predicted numbers of signal and background events, as well as the total yield, together with the observed number of events in data, are shown after the combined maximum-likelihood fit for the $\Pe$+jets (top) and $\mu$+jets (bottom) categories. The uncertainties include the statistical component from the fit, but not the theoretical uncertainties in the $\ttbar$ signal. The uncertainties in the sum of backgrounds and the total yield are determined neglecting correlations for presentational purposes, although the full likelihood with correlations is used to compute the uncertainties in the measurements of the cross section.}
\centering
\begin{scotch}{lxxx}
\multirow{2}{*}{Sample} & \multicolumn{3}{c}{Number of events ($\Pe$+jets)} \\
 & \multicolumn{1}{c}{0t} & \multicolumn{1}{c}{1t+0b} & \multicolumn{1}{c}{1t+1b} \\
\hline
\ttbar signal      & 1560 , 120 & 289 , 22 & 226 , 17 \\
\ttbar other & 458 , 34 & 40.0 , 3.0 & 30.1 , 2.3 \\
Single \PQt  & 260 , 120 & 11.6 , 5.3 & 3.2 , 1.5 \\
$\PW$+jets               & 3670 , 250 & 130 , 9 & 2.7 , 0.2 \\
Multijet                  & 760 , 130 & 68 , 11 & 10.5 , 1.8 \\
\hline
Total background & 5140 , 310 & 249 , 16 & 46.5 , 3.2 \\
Signal + background  & 6700 , 330 & 537 , 27 & 273 , 17 \\
\hline
Data                 & \multicolumn{1}{c}{6833} & \multicolumn{1}{c}{538} & \multicolumn{1}{c}{242} \\
\end{scotch}
\null\vspace*{1em}
\begin{scotch}{lxxx}
\multirow{2}{*}{Sample} & \multicolumn{3}{c}{Number of events ($\mu$+jets)} \\
 & \multicolumn{1}{c}{0t} & \multicolumn{1}{c}{1t+0b} & \multicolumn{1}{c}{1t+1b} \\
\hline
\ttbar signal     & 1920 , 140 & 359 , 27 & 271 , 20 \\
\ttbar other & 478 , 36 & 44.7 , 3.4 & 29.7 , 2.2 \\
Single \PQt & 290 , 140 & 14.4 , 6.6 & 4.1 , 1.9 \\
$\PW$+jets               & 4790 , 330 & 154 , 11 & 3.9 , 0.3 \\
Multijet                  & 360 , 170 & 13.4 , 6.3 & 7.6 , 3.6 \\
\hline
Total background & 5920 , 390 & 226 , 14 & 45.3 , 4.6 \\
Signal + background  & 7840 , 420 & 586 , 31 & 317 , 21 \\
\hline
Data                 & \multicolumn{1}{c}{7712} & \multicolumn{1}{c}{622} & \multicolumn{1}{c}{306} \\
\end{scotch}
\end{table*}

\begin{figure*}[htbp]
\centering
\includegraphics[width=0.45\textwidth]{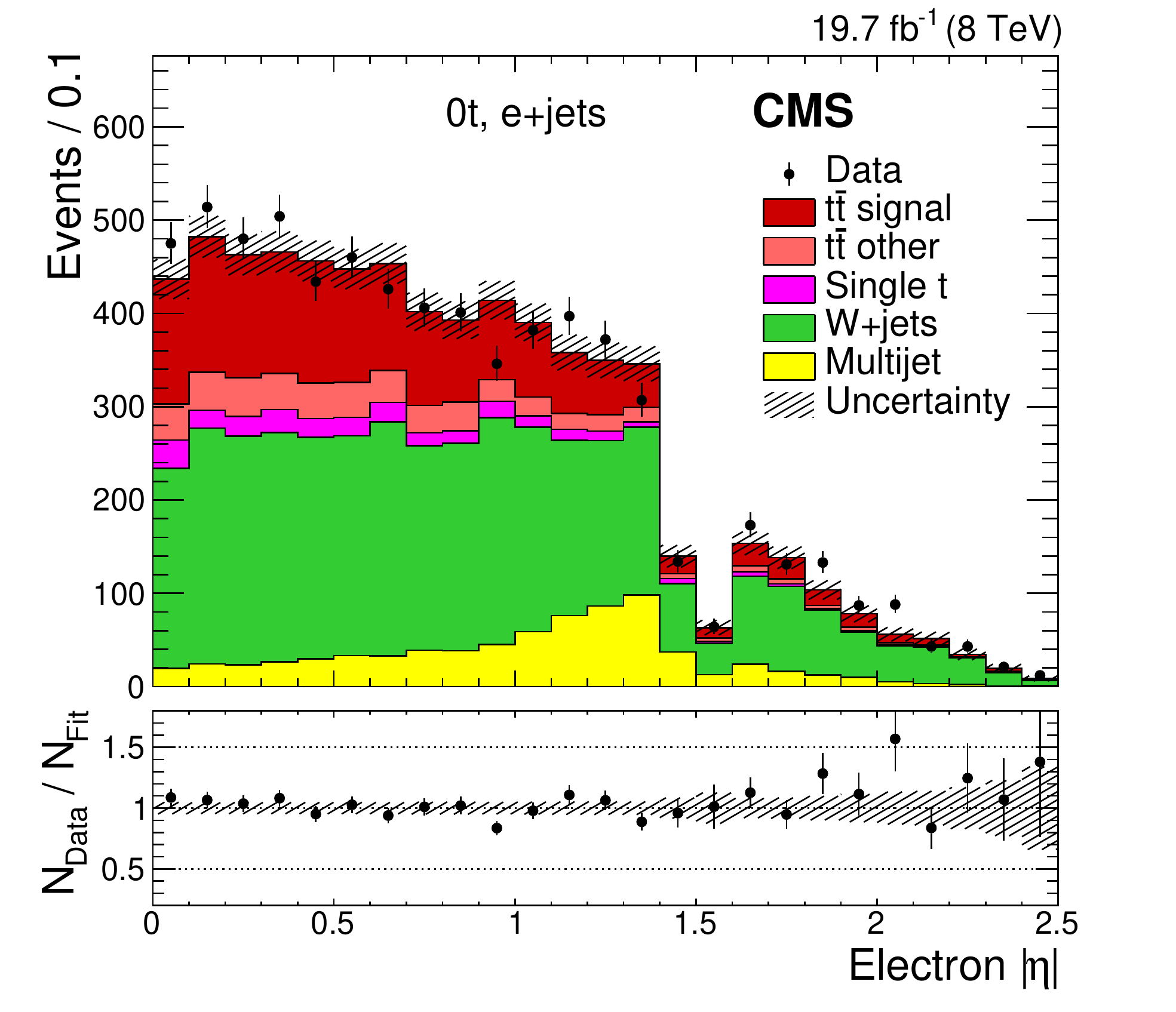}
\includegraphics[width=0.45\textwidth]{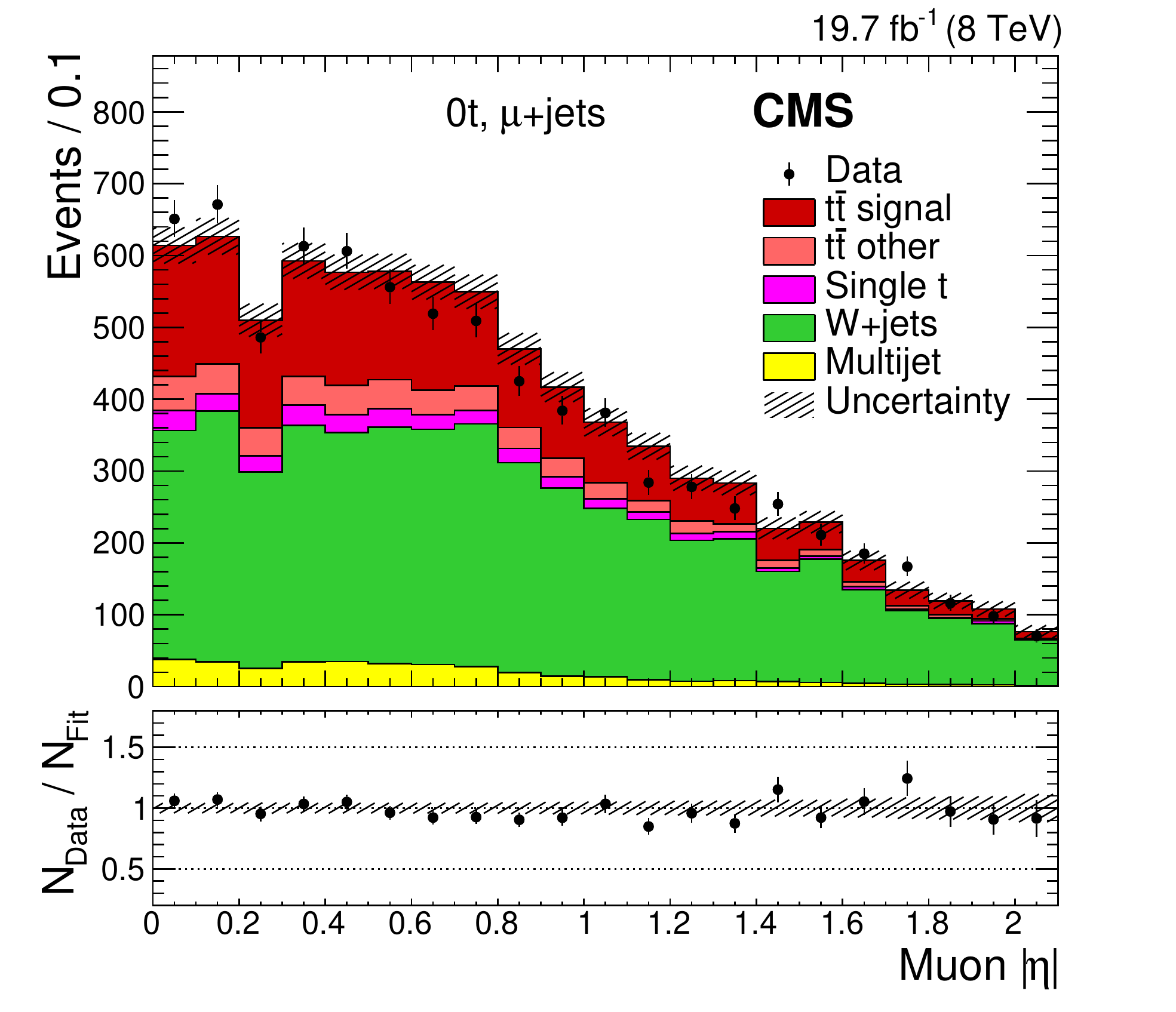}\\
\includegraphics[width=0.45\textwidth]{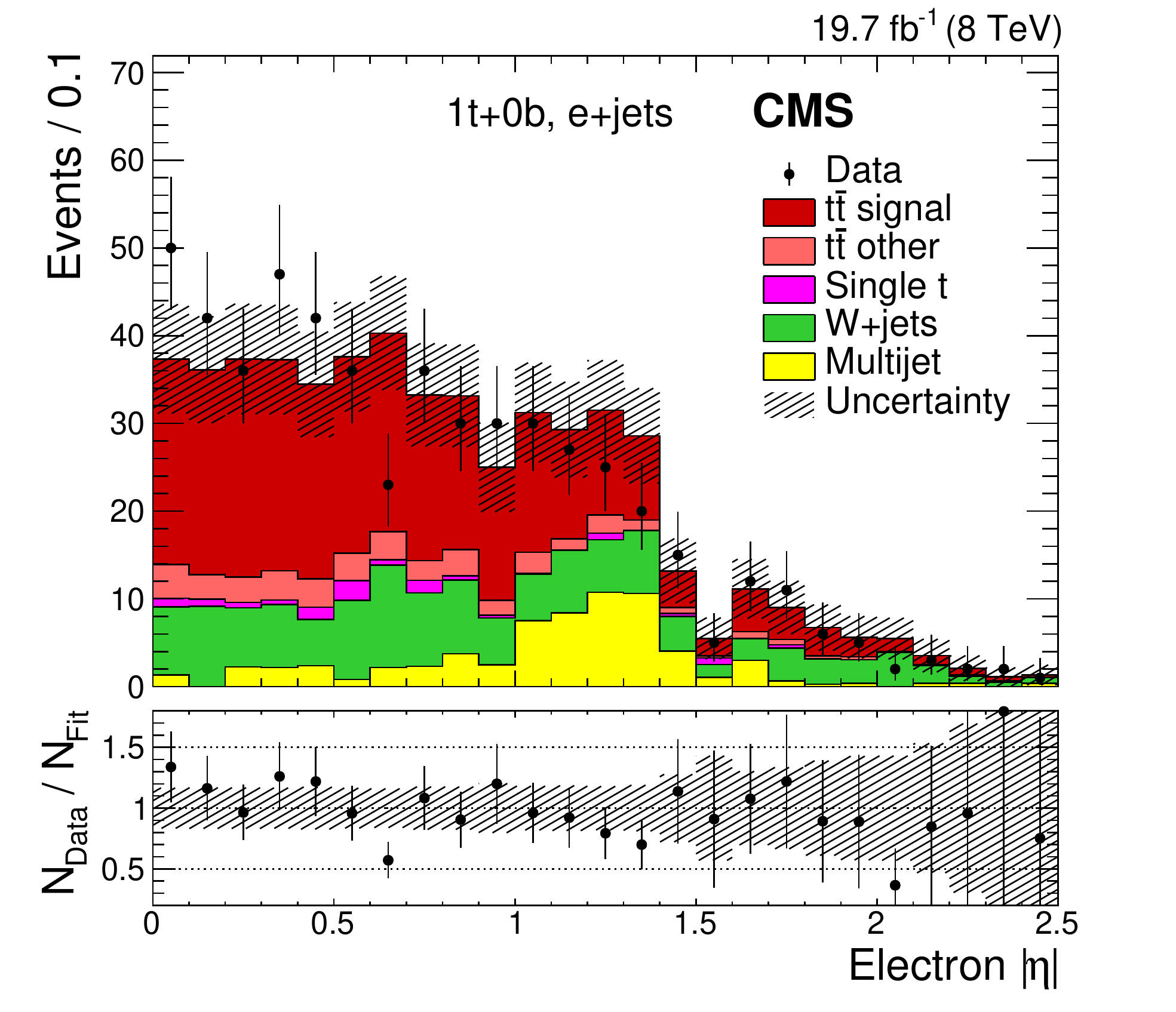}
\includegraphics[width=0.45\textwidth]{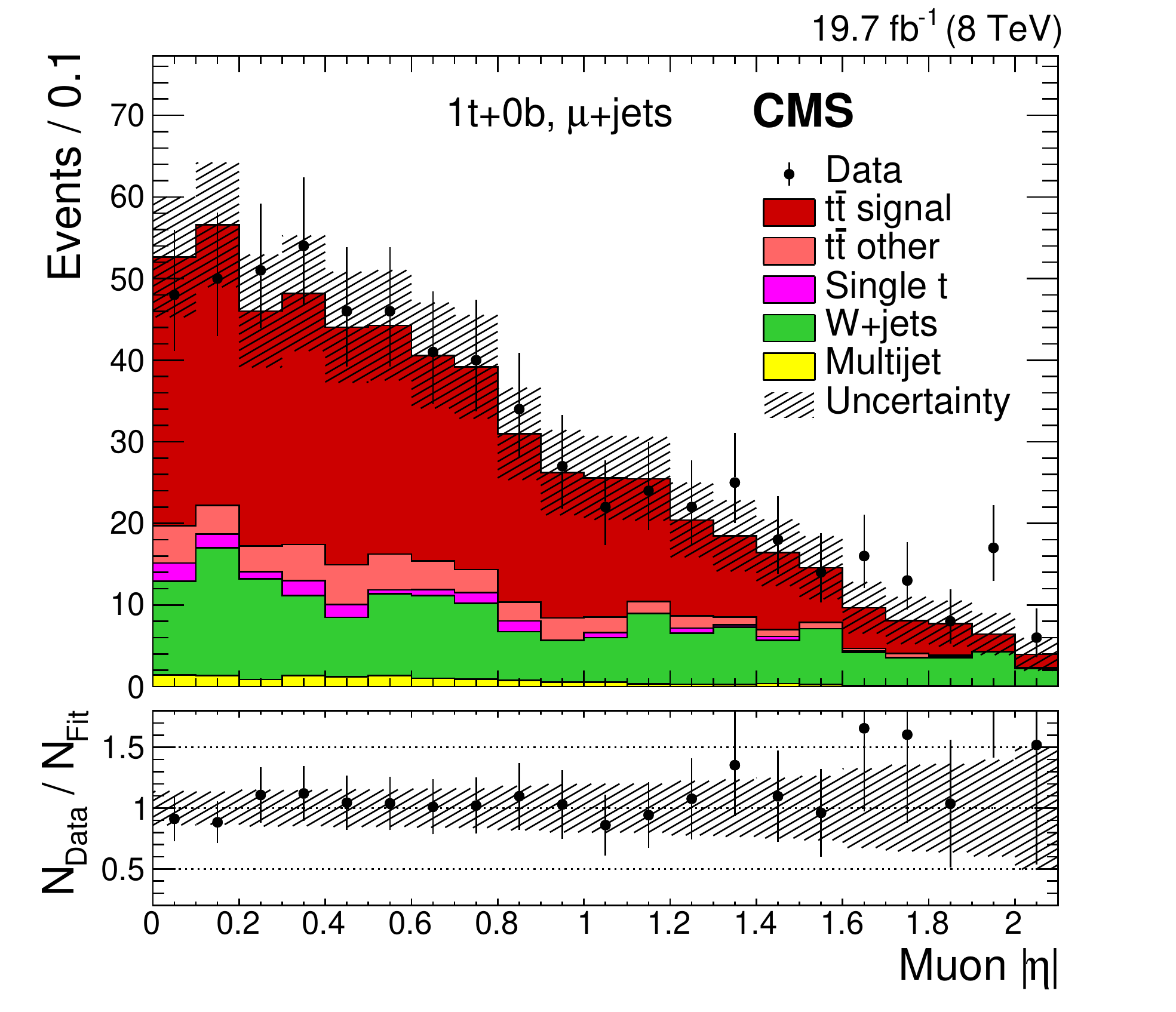}\\
\includegraphics[width=0.45\textwidth]{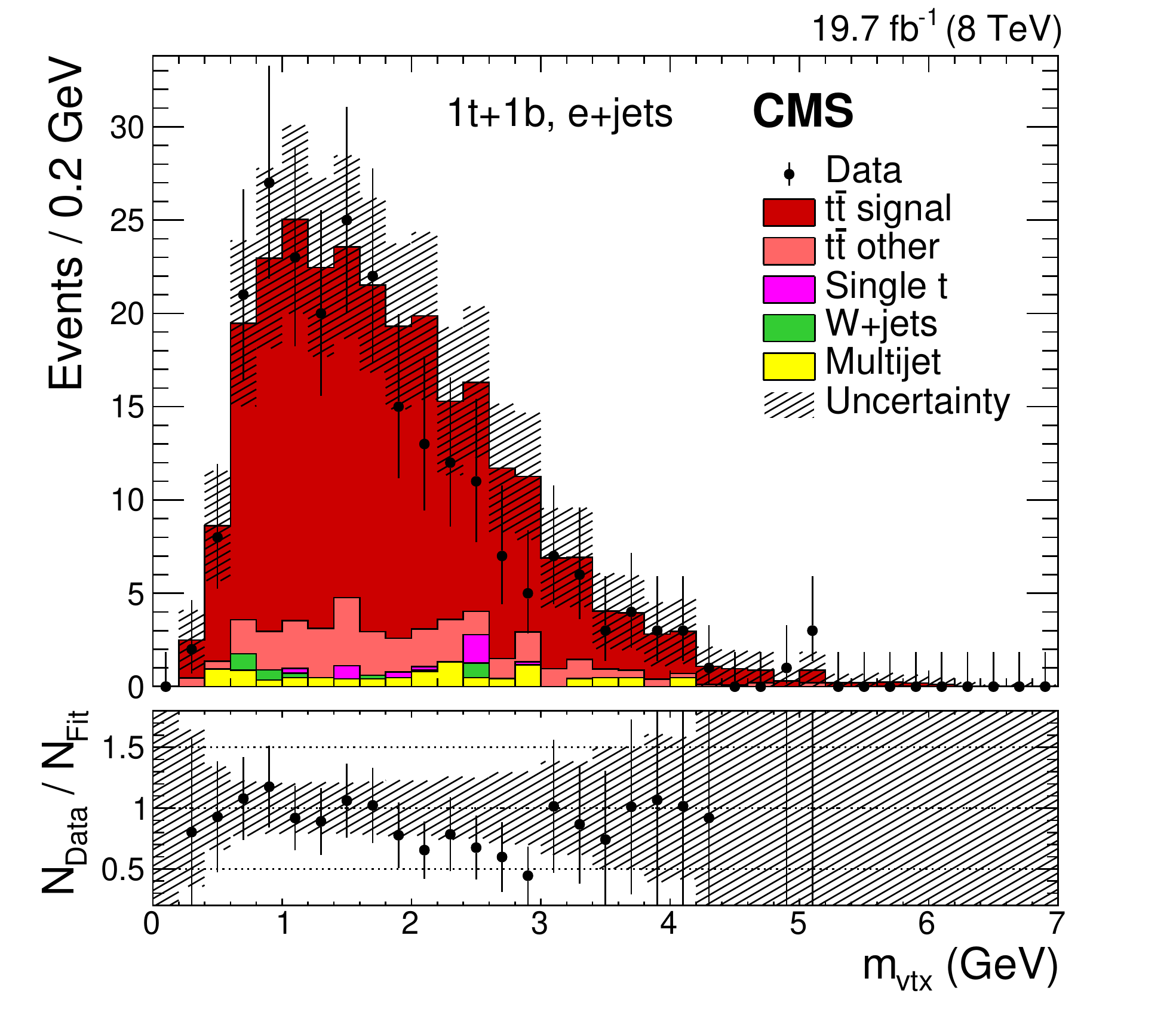}
\includegraphics[width=0.45\textwidth]{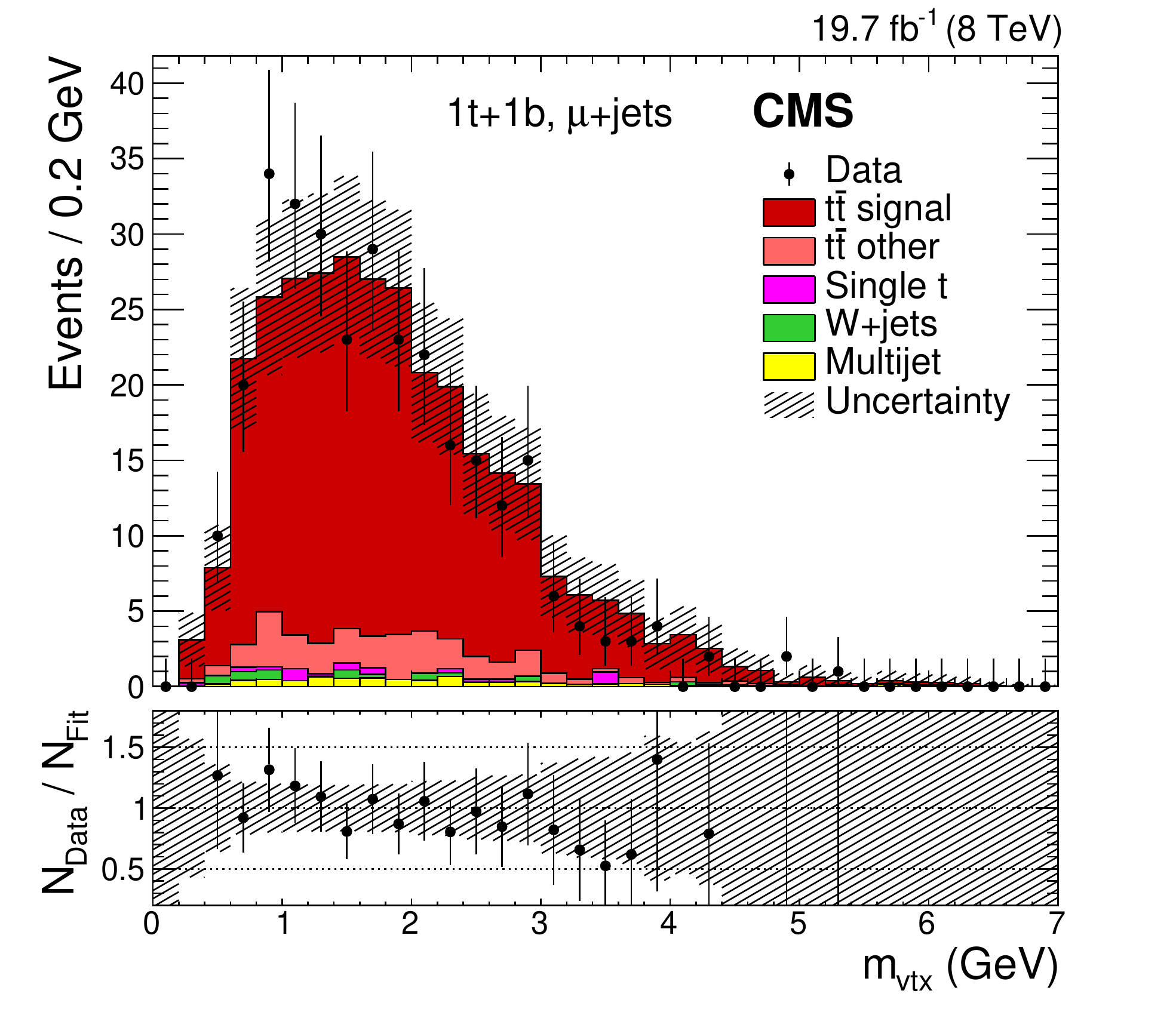}
\caption{Lepton $\abs{\eta}$ and $m_\text{vtx}$ distributions from data (points) and for signal and background sources (histograms) with normalizations from the fit for the 0t (top), 1t+0b (middle), and 1t+1b (bottom) event categories, for the $\Pe$+jets (left column) and $\mu$+jets (right column) channels. The vertical bars on the data points represent the statistical uncertainties. The shaded bands reflect the combined statistical and experimental systematic uncertainties after the fit to the signal and background yields, added in quadrature neglecting their correlations for presentational purposes. The ratios of data ($N_\text{Data}$) to the total prediction from the fit ($N_\text{Fit}$) are shown below each panel, along with the uncertainty band from the fit.
\label{fig:posterior_values_comb}}
\end{figure*}

\begin{figure*}[htbp]
\centering
\includegraphics[width=0.45\textwidth]{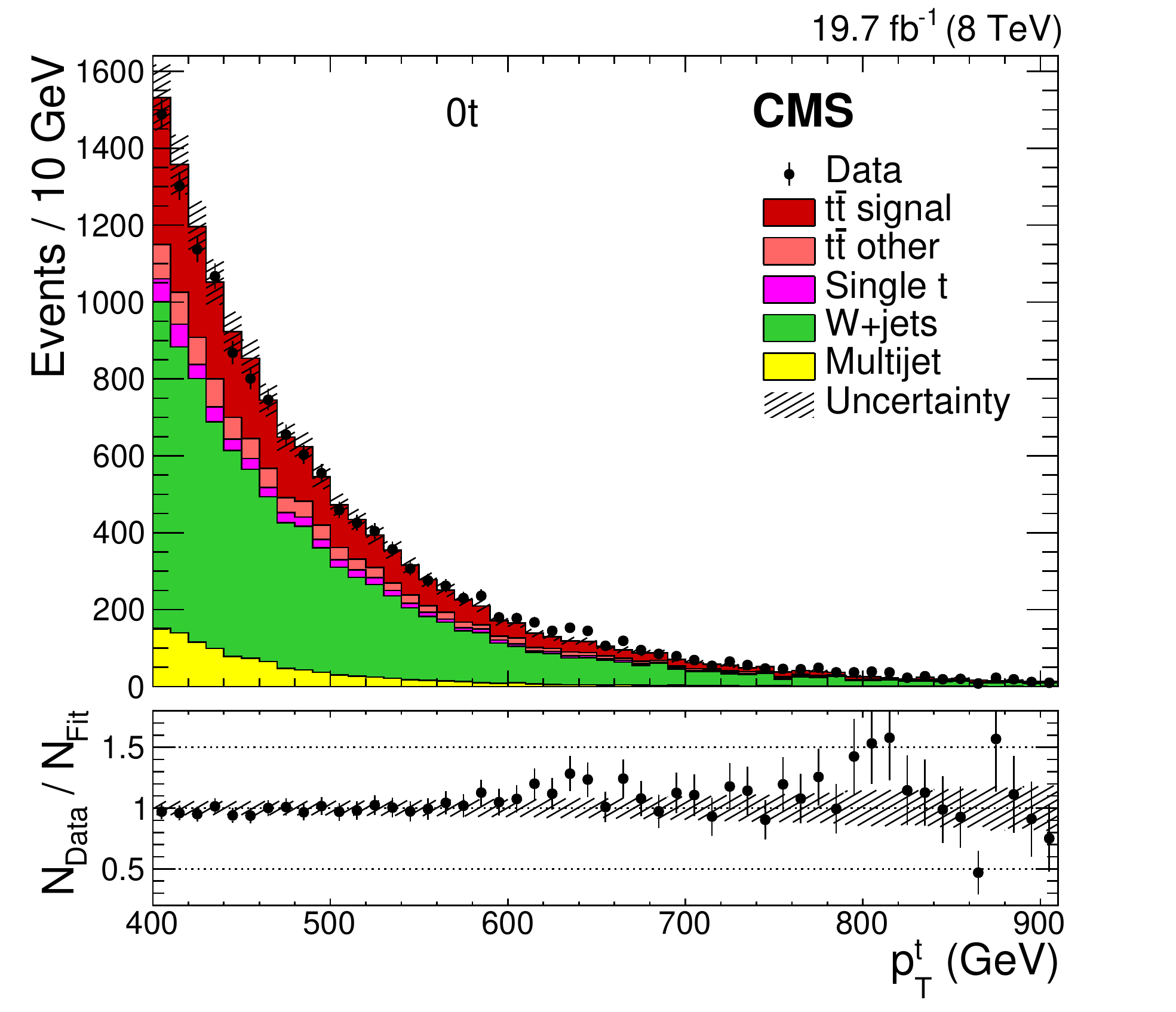}
\includegraphics[width=0.45\textwidth]{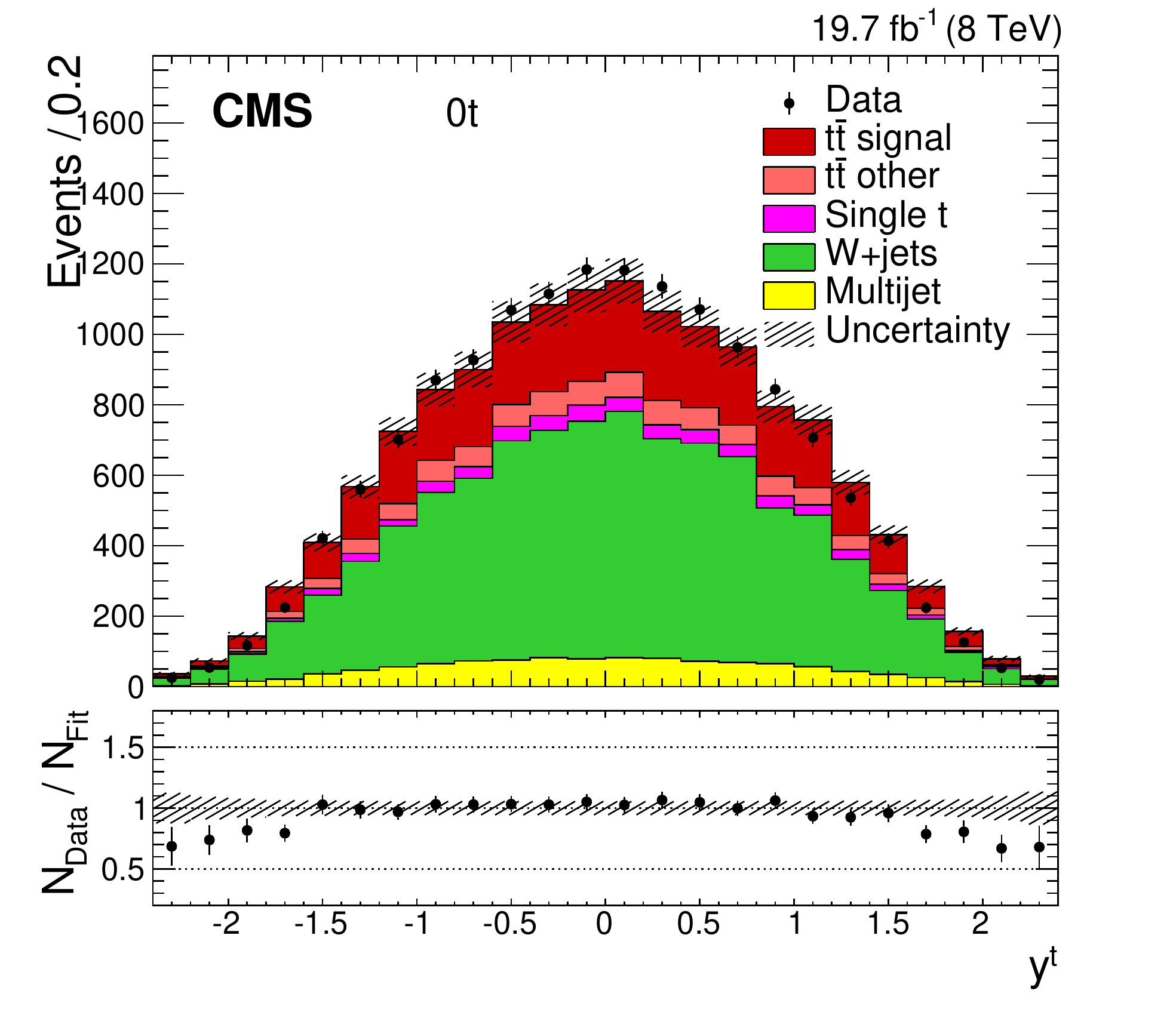}\\
\includegraphics[width=0.45\textwidth]{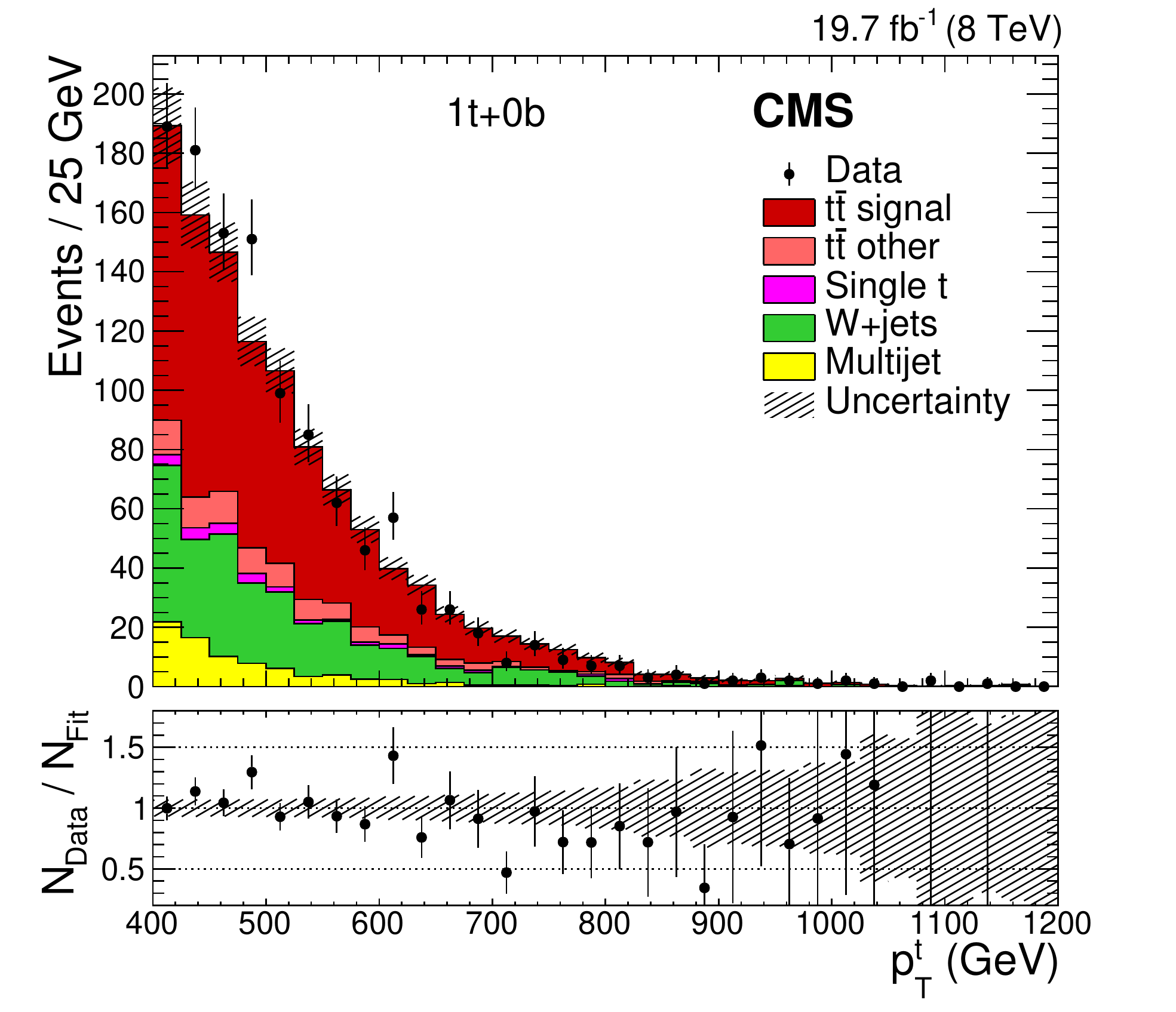}
\includegraphics[width=0.45\textwidth]{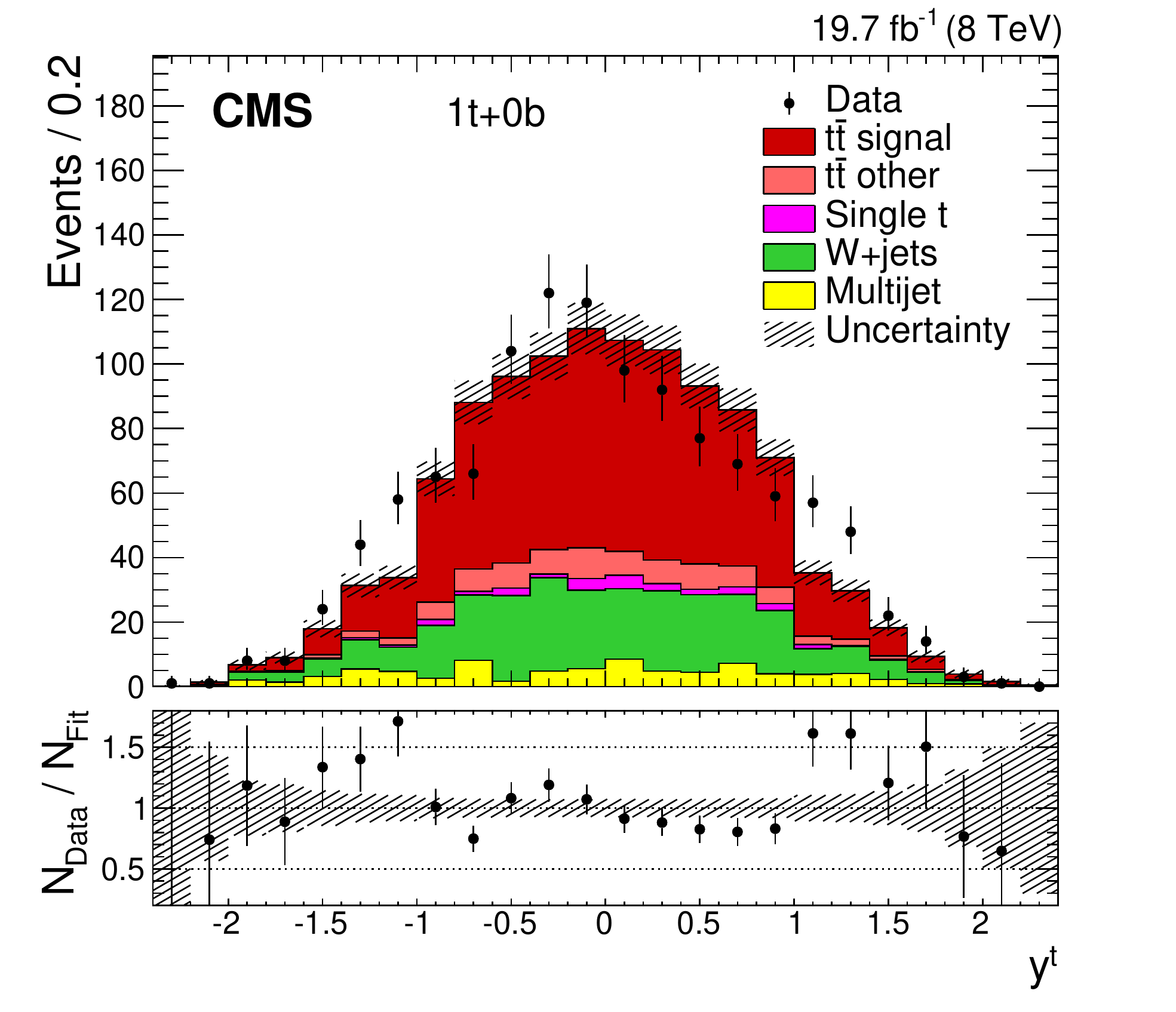}\\
\includegraphics[width=0.45\textwidth]{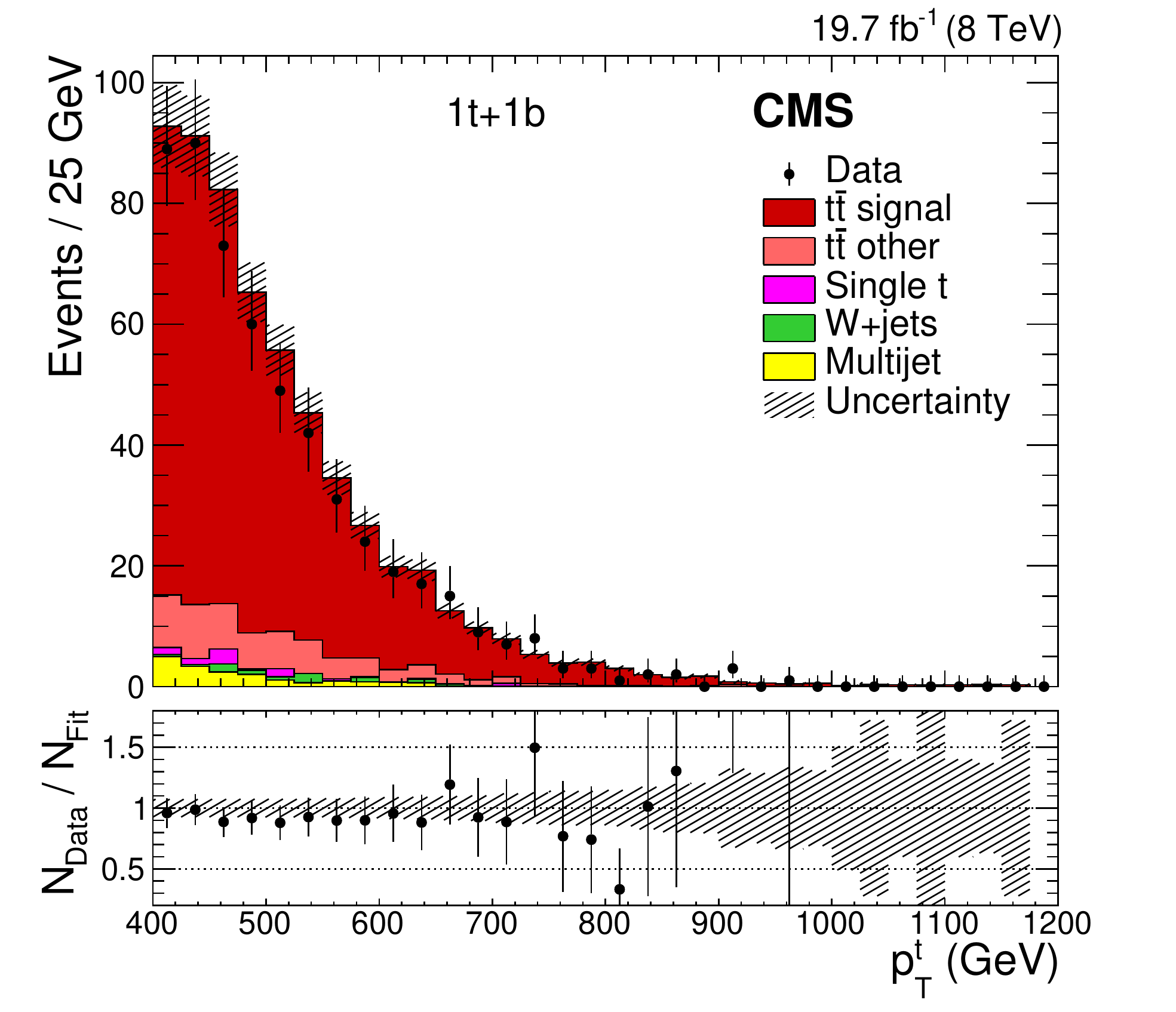}
\includegraphics[width=0.45\textwidth]{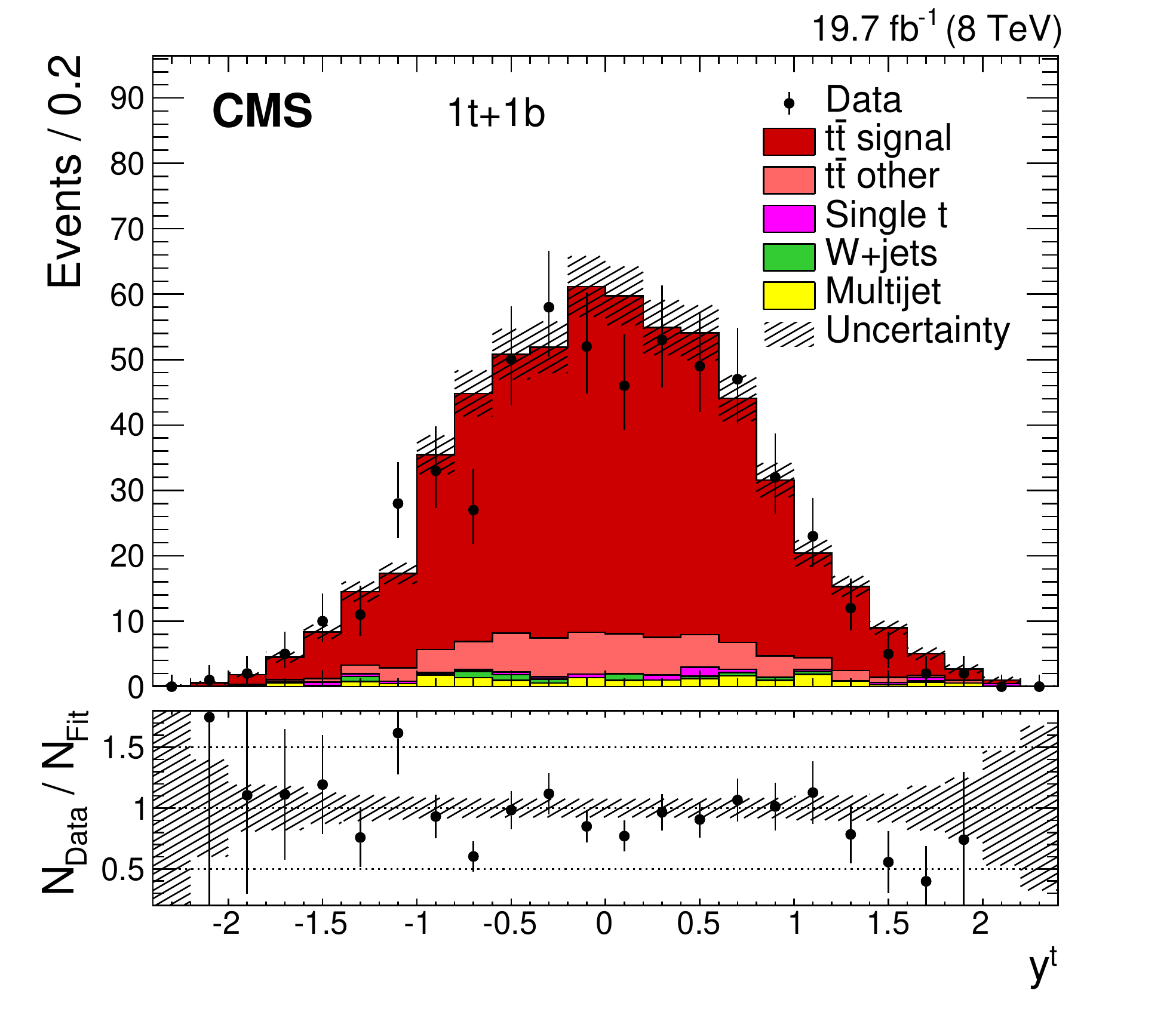}
\caption{Transverse momentum (left column) and rapidity (right column) distributions of the hadronic-side, large-$R$ jet for the 0t (top), 1t+0b (middle), and 1t+1b (bottom) event categories, combining the $\Pe$+jets and $\mu$+jets channels. The data are compared to the total signal and background yields using normalizations from the maximum-likelihood fit. The vertical bars on the data points represent the statistical uncertainties. The shaded bands reflect the combination of the statistical and post-fit systematic uncertainties in the signal and background yields added in quadrature, without the uncertainties based on the form of the distributions, and neglecting their correlations for presentational purposes. The ratios of data ($N_\text{Data}$) to the total prediction from the fit ($N_\text{Fit}$) are shown below each panel, along with the uncertainty band from the fit.
\label{fig:posterior_toppt_y_comb}}
\end{figure*}

The distributions in $\abs{\eta}$ and $m_{\text{vtx}}$ after the combined maximum-likelihood fit to $\Pe$+jets and $\mu$+jets events are shown in Fig.~\ref{fig:posterior_values_comb}, comparing the fitted values of the model to the data from each of the fitted categories (0t, 1t+0b, 1t+1b). The uncertainty bands show the combined fitted statistical and experimental systematic uncertainties in the signal and backgrounds, added in quadrature neglecting correlations for presentational purposes, although the full likelihood with correlations is used to compute the uncertainties in the measurements of the cross section.
The $\pt$ and $y$ distributions of the hadronic-side, large-$R$ jet are shown for each category in Fig.~\ref{fig:posterior_toppt_y_comb}.
These figures show the data, together with the signal and background yields from simulation (or, for multijet background, from data enhanced with multijet events), using the normalizations from the fit, as well as the ratio of the data to the total fit.
Since the $\pt^{\PQt}$ and $y^{\PQt}$ variables are not used in the fit, the signal and background distributions in Fig.~\ref{fig:posterior_toppt_y_comb} are taken from simulation (or the data sideband for the multijet background). In extracting the differential cross sections, these distributions are used for the backgrounds, while the signal is taken from the data after subtracting the background contributions.

\subsection{Integrated \texorpdfstring{$\ttbar$}{t-tbar} cross section measurement}
\label{sec:xsec_int}

The measurement at the particle level is defined within a fiducial region designed to closely match the event selections in the detector and minimize the dependence on theoretical input. The measurement at the parton level is defined relative to the top and antitop quarks before they decay, but after they radiate any gluons.

The {\PPsix} simulation is used to determine the acceptance for the particle-level and parton-level selections and to obtain the predicted cross section values. The following particle-level selections are used to define the fiducial region in the simulation:
\begin{enumerate}
\renewcommand{\labelenumi}{(\roman{enumi})}
\item One electron or muon with $\pt > 45\GeV$ (computed prior to any potential photon radiation) and $\abs{\eta} < 2.1$.
\item At least one anti-$\kt$ ($R=0.5$) jet with $0.1 < \Delta R(\ell, \text{jet}) < \pi/2$, $\pt > 30\GeV$, and $\abs{\eta}<2.4$.
\item At least one CA ($R=0.8$) jet with $\Delta R(\ell, \text{jet}) > \pi/2$, $\pt > 400\GeV$, $140 < m_{\Pj} < 250\GeV$, and $\abs{\eta}<2.4$.
\end{enumerate}

Jets at the particle level in the simulation are formed from stable particles, excluding electrons, muons, and neutrinos. The cross section at parton level is measured for the region where the top or antitop quark that decays to quarks has $\pt > 400\GeV$. No other kinematic requirements are imposed.

The measurements at both the particle and parton levels are corrected for the branching fraction of $\ttbar \to \Pe/\mu$+jets, determined from the $\ttbar$ simulation.

The integrated $\ttbar$ cross section is obtained from the $\ttbar$ signal yield in the maximum-likelihood fit.
Uncertainties associated with the signal modeling are not included as nuisance parameters in the fit. These are instead evaluated through the difference in the signal acceptance from changes made in the $\mu_R$ and $\mu_F$ scales and PDF variations. The uncertainties from the choice of event generator and parton shower algorithm are also evaluated independently of the fit through the difference in the $\ttbar$ signal acceptance between the {\PPsix} and {\MCATHsix} predictions at the particle and parton levels.

The measurements of the integrated cross sections for $\pt^{\PQt} > 400\GeV$ are:
\ifthenelse{\boolean{cms@external}}{
\begin{equation*}
\begin{aligned}
{\text{particle level: }}
\sigma_{\ttbar} = & \ 0.499 \pm 0.035\,\text{(stat+syst)} \\
& \pm 0.095\thy \pm 0.013\lum\unit{pb},\\
{\text{parton level: }}
\sigma_{\ttbar} = & \ 1.44 \pm 0.10\,\text{(stat+syst)} \\
& \pm 0.29\thy \pm 0.04\lum\unit{pb}.
\end{aligned}
\end{equation*}
}
{
\begin{equation*}
\begin{aligned}
 \text{particle level: }  \sigma_{\ttbar} = & 0.499 \pm 0.035\,\text{(stat+syst)} \pm 0.095\thy \pm 0.013\lum\unit{pb},\\
 \text{parton level: }  \sigma_{\ttbar} = & 1.44 \pm 0.10\,\text{(stat+syst)} \pm 0.29\thy \pm 0.04\lum\unit{pb}.
\end{aligned}
\end{equation*}
}

The theoretical uncertainties from the PDF, $\mu_R$ and $\mu_F$ scales, and choice of event generator and parton shower algorithm are, respectively, 9\%, 9\%, and 14\% at the particle level, and 9\%, 10\%, and 15\% at the parton level.

The measurements are compared to predictions from different $\ttbar$ simulations. Assuming the NNLO cross section of 252.9\unit{pb}~\cite{topNNLO} for the full phase space, the resulting {\PPsix} cross section is 0.580 (1.67)\unit{pb} at particle (parton) level.
The ratio of the measured integrated $\ttbar$ cross section for the high-$\pt$ region to the value predicted by the {\PPsix} simulation is $0.86 \pm 0.16$ ($0.86 \pm 0.19$) for the particle (parton) level.
Thus, the measurements and predictions are consistent within the total uncertainty, which is dominated by the theoretical uncertainty in the cross section extraction.
The integrated cross sections are also extracted from the {\MADPsix} and {\MCATHsix} simulations, again assuming the NNLO cross section for the full phase space, and are 0.675 (1.85)\unit{pb} and 0.499 (1.42)\unit{pb} at the particle (parton) level, respectively. The prediction from the {\MCATHsix} simulation agrees well with the measured values, while the {\MADPsix} simulation overestimates the cross sections at both particle and parton levels.

\subsection{Differential \texorpdfstring{$\ttbar$}{t-tbar} cross section measurements}
\label{sec:xsec_diff}

The differential $\ttbar$ cross section is measured as a function of the $\pt$ and $y$ of the top quark that decays to a hadronic final state.
The event sample from which the $\pt$ and $y$ distributions of the $\PQt$ jet candidates are extracted is defined by combining the signal-dominated 1t+0b and 1t+1b event categories.
The observed number of $\ttbar$ events at detector level is first extracted from data by subtracting the SM background contributions using the normalizations from the maximum-likelihood fit (shown in Table~\ref{tab:counts_postfit}). As a cross-check, it is verified that a small $\ttbar$ contribution added to the maximum-likelihood fit from a beyond-the-SM process, such as a 1--2\% contribution from $\Z' \to \ttbar$ (corresponding to a signal cross section already excluded in Ref.~\cite{B2G-13-008}), has a negligible impact on the extracted SM backgrounds.
We also verify that a small potential modification of the top quark rapidity has a minimal impact on the background normalizations that is well within the quoted background normalization uncertainties. 

An unfolding procedure translates the observed number of $\ttbar$ events in bins of reconstructed $\pt$ and $y$ of the $\PQt$ jet candidate to a cross section in bins of particle- and parton-level top quark $\pt^{\PQt}$ and $y^{\PQt}$. If more than large-$R$ jet fulfills the particle-level selection in Section~\ref{sec:xsec_int}, which occurs for $<$1\% of events, the one with highest $\pt$ is chosen as the particle-level \PQt jet.
The unfolding accounts for all reconstruction and detector efficiencies, detector resolution effects, and migrations of $\ttbar$ events across bins. The unfolding is performed using response matrices, determined with simulated {\PPsix} $\ttbar$ events, using the singular-value-decomposition (SVD) method~\cite{Hocker:1995kb} in the \textsc{RooUnfold} package~\cite{RooUnfold}. 

The background-subtracted data are unfolded in two steps, first from detector level to particle level, and in a second step from particle level to parton level. Response matrices are created between the $\pt$ and $y$ of the reconstructed $\PQt$ jet candidate and the particle-level $\PQt$ jet, and between the particle-level $\PQt$ jet and the parton-level top quark. These response matrices are used to unfold the data and obtain the differential cross sections, after dividing by the bin width and correcting for the branching fraction of $\ttbar \to \Pe/\mu$+jets. The unfolding is performed multiple times, repeating the procedure for each systematic change that affects the $\pt^{\PQt}$ or $y^{\PQt}$ distributions. 
The electron and muon channels are unfolded separately, and are then combined through the statistically weighted mean in each bin. Specifically, the combined cross section in a bin ($\sigma$) is given by $\sigma = \sum(\sigma_i / \delta \sigma_i^2) / \sum(1 / \delta \sigma_i^2)$, where $\sigma_i$ is the cross section in a bin for each channel ($i=\Pe, \mu$) and $\delta \sigma_i$ is the corresponding uncertainty. The statistical uncertainty in the combined cross section ($\delta \sigma$) is given by $\delta \sigma = 1 / (\sum(1 / \delta \sigma_i^2))^{1/2}$. The combination is repeated for each systematic variation, and the difference with respect to the combined nominal value is taken as the uncertainty for that source of systematic bias. 
The uncertainty in the normalization of the background is extracted by rescaling the subtracted background by $\pm1$ standard deviation, as derived from the maximum-likelihood fit in Section~\ref{sec:xsec_fit}, and taking the difference in the unfolded result relative to the nominal yield as the uncertainty at particle and parton level, respectively. Similarly, the t tagging efficiency uncertainty as measured at detector level is translated into an uncertainty in the differential measurement at particle and parton levels by unfolding, assuming systematically varied t tagging efficiencies. 
The uncertainties from the choice of event generator and parton shower algorithm are evaluated by unfolding the nominal {\PPsix} simulated events using the response matrix from {\MCATHsix}. The differences between the unfolded simulation and the predictions at the particle and parton levels are taken as the uncertainties.  
At particle (parton) level, these are 1--18\% (2--21\%) and 3--8\% (2--6\%) for the $\pt^{\PQt}$ and $y^{\PQt}$ measurements, respectively.

\begin{figure*}[htbp]
\centering
\includegraphics[width=0.47\textwidth]{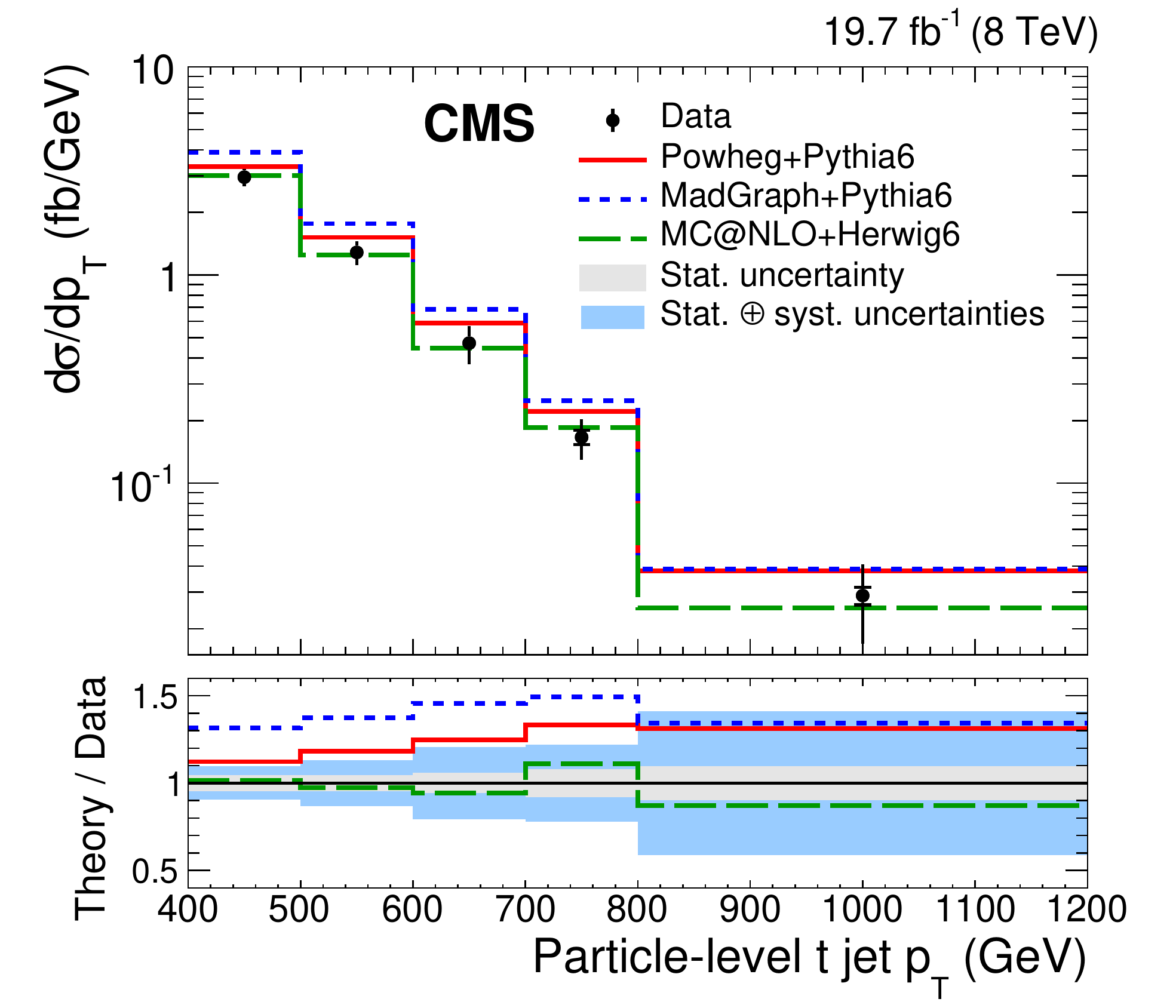}
\includegraphics[width=0.47\textwidth]{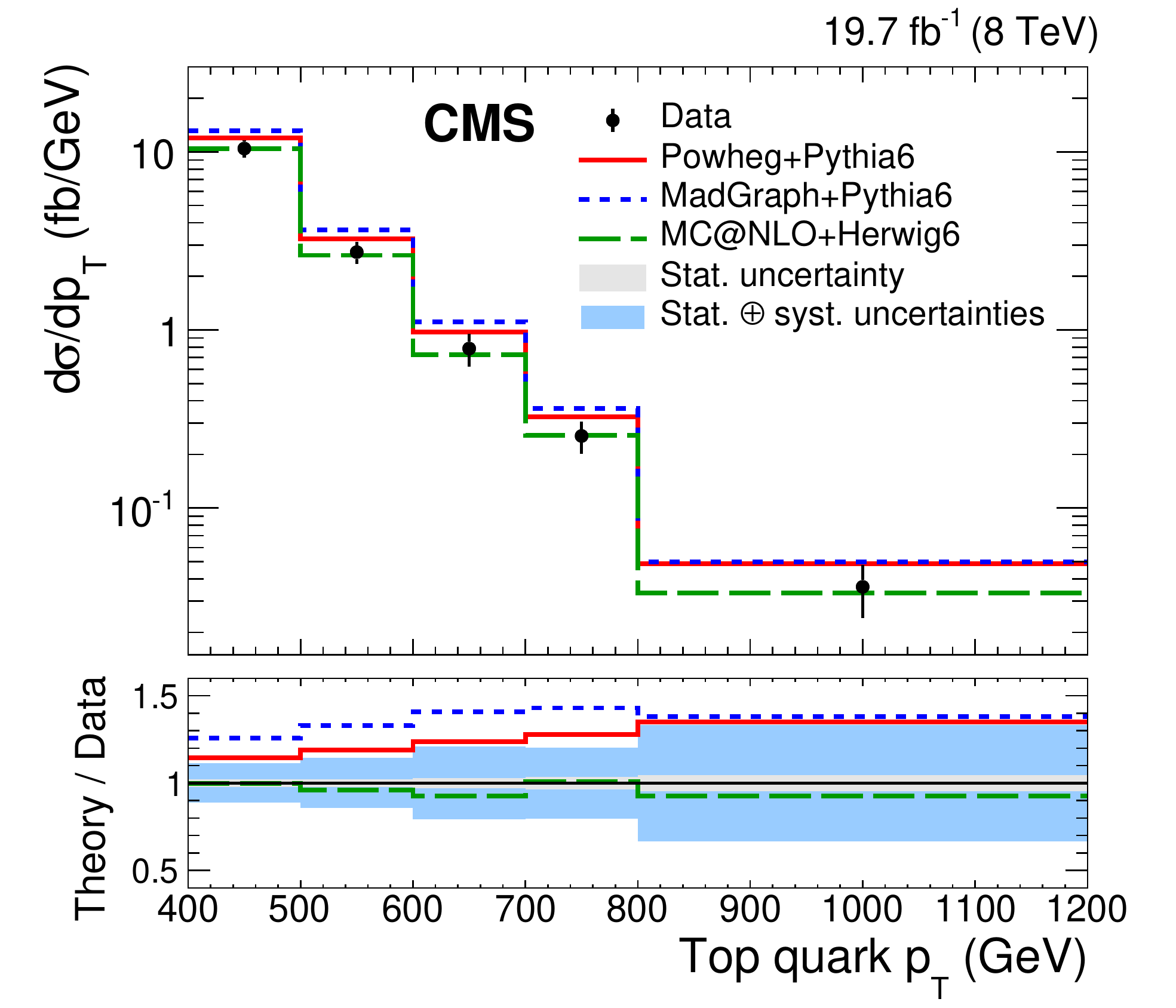} \\
\includegraphics[width=0.47\textwidth]{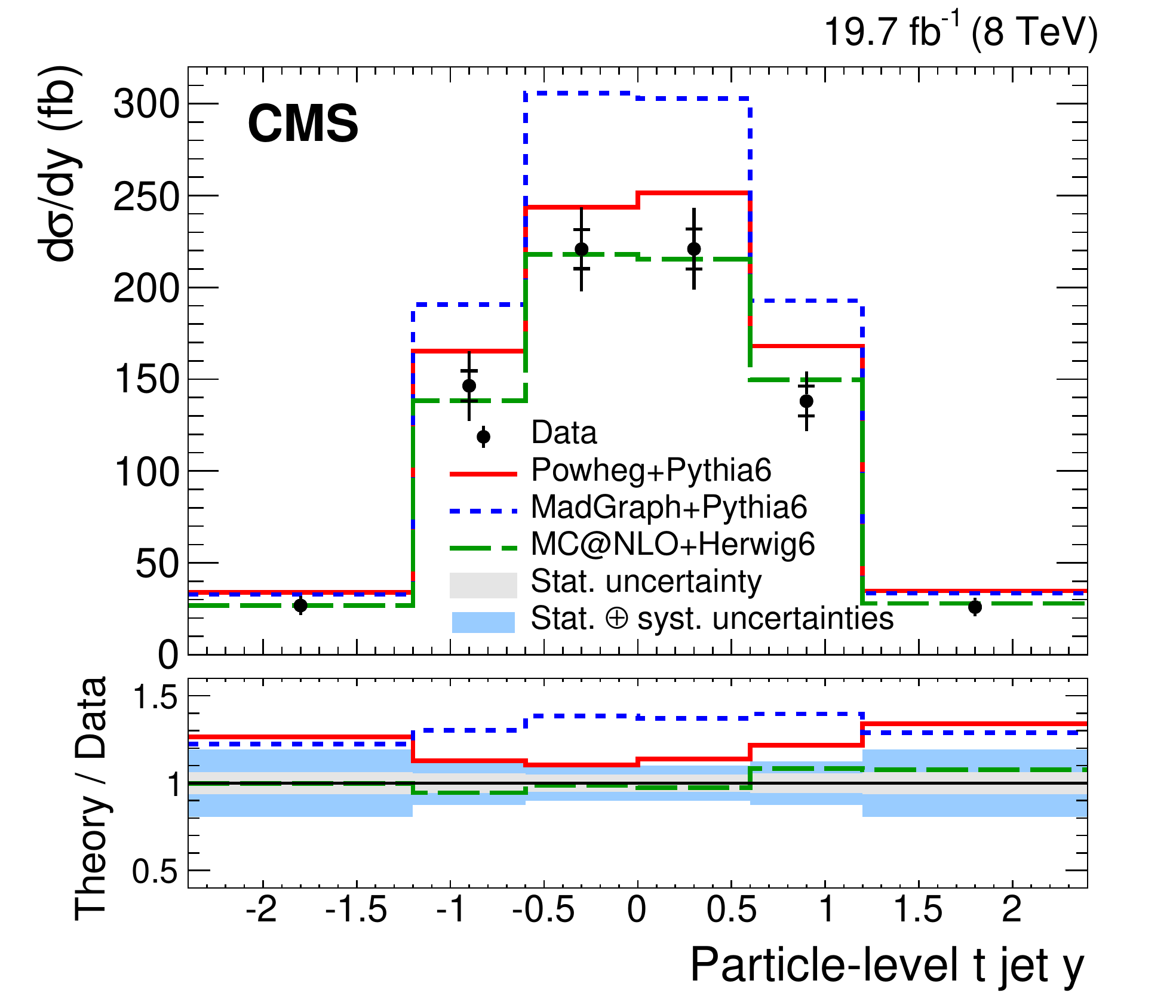}
\includegraphics[width=0.47\textwidth]{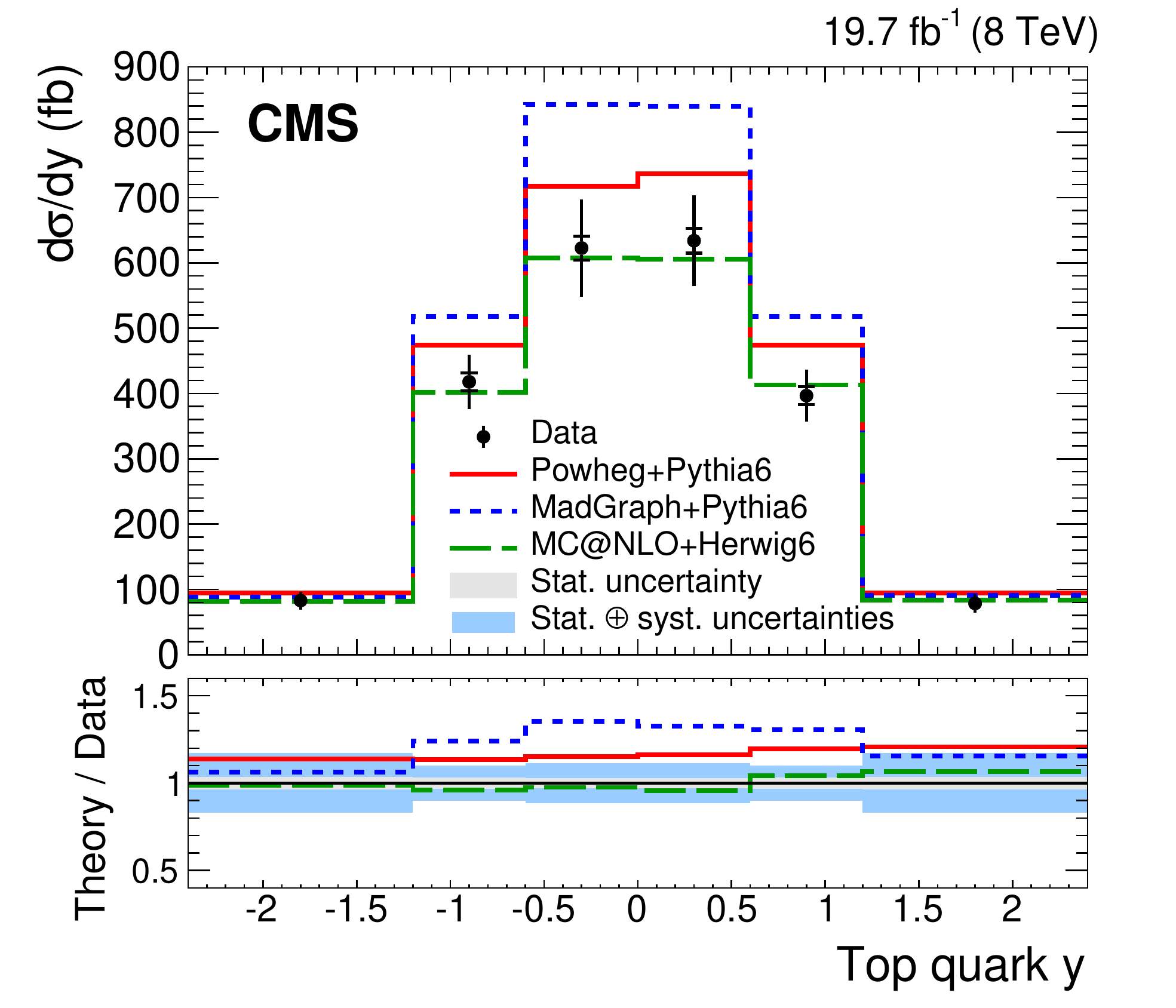}
\caption{\label{fig:unfoldWithError_comb} Differential $\ttbar$ cross section in bins of particle-level $\PQt$ jet $\pt$ (top left), parton-level top quark $\pt$ (top right), particle-level $\PQt$ jet $y$ (bottom left), and parton-level top quark $y$ (bottom right), including all systematic uncertainties. The lower plots show the ratio of the theoretical predictions to the data. The statistical uncertainties are represented by the inner vertical bars with ticks and the light bands in the ratios. The combined uncertainties are shown as full vertical bars and the dark solid bands in the ratios.}
\end{figure*}

\begin{figure*}[htbp]
\centering
\includegraphics[width=0.47\textwidth]{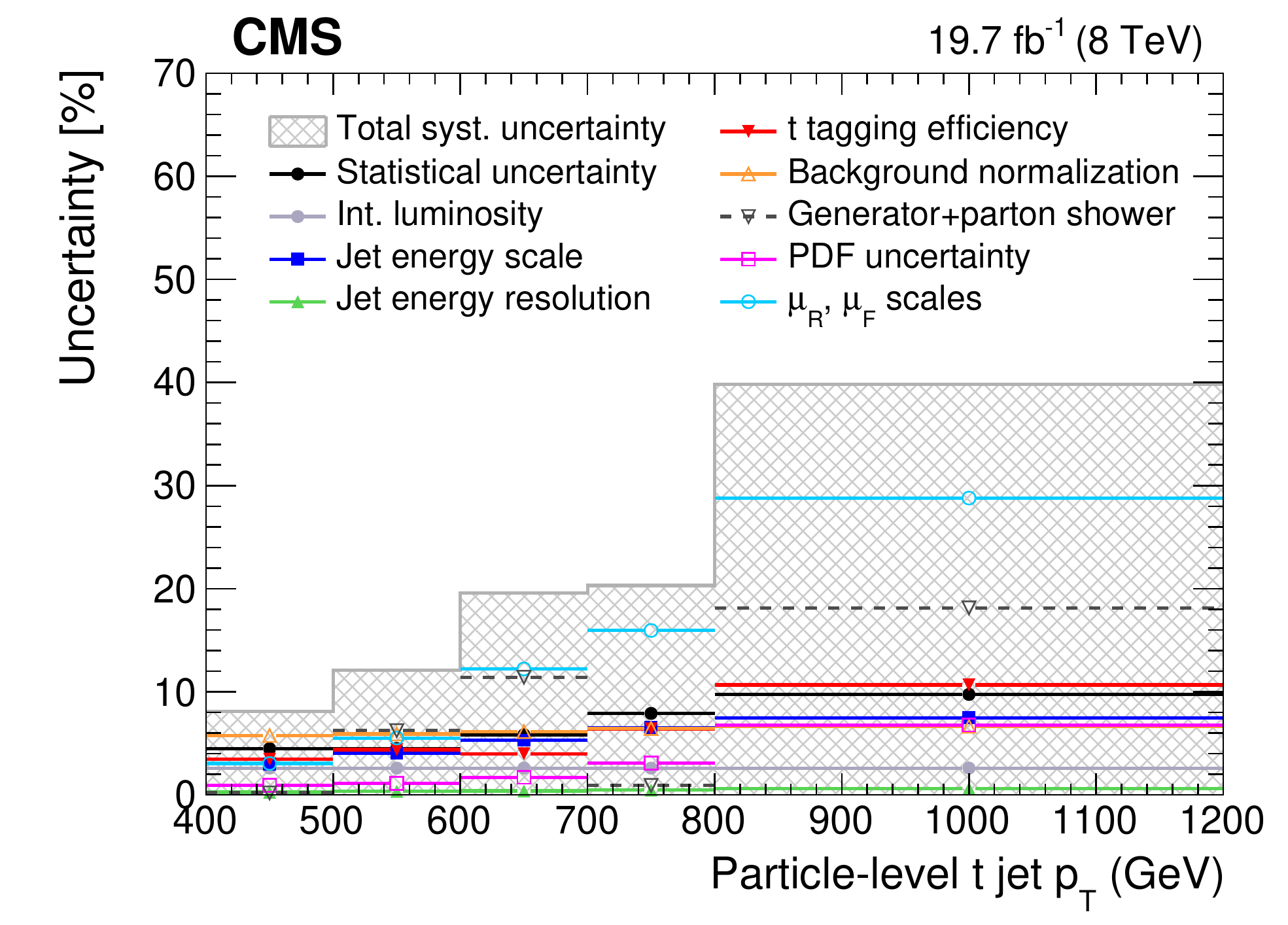}
\includegraphics[width=0.47\textwidth]{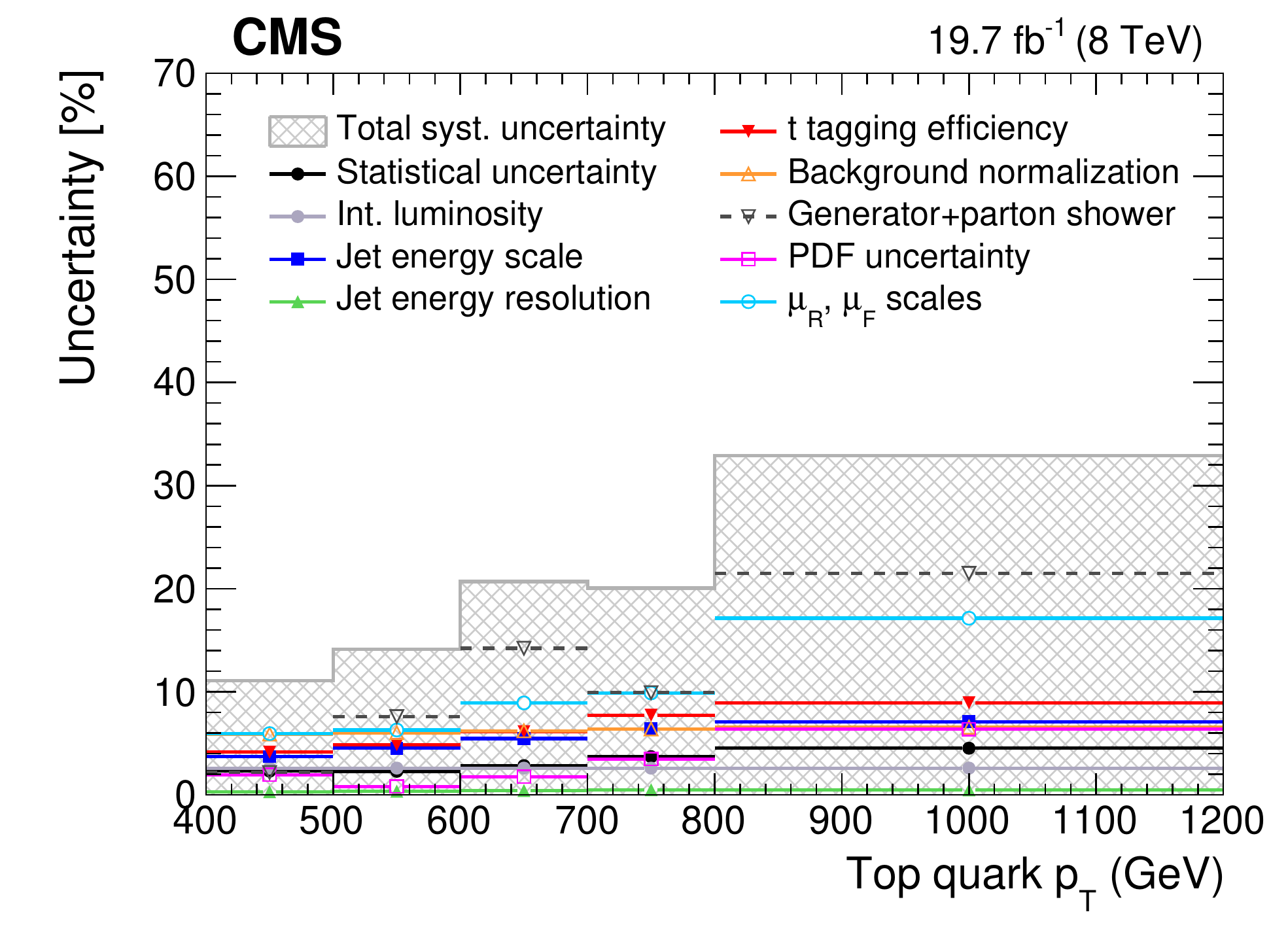} \\
\includegraphics[width=0.47\textwidth]{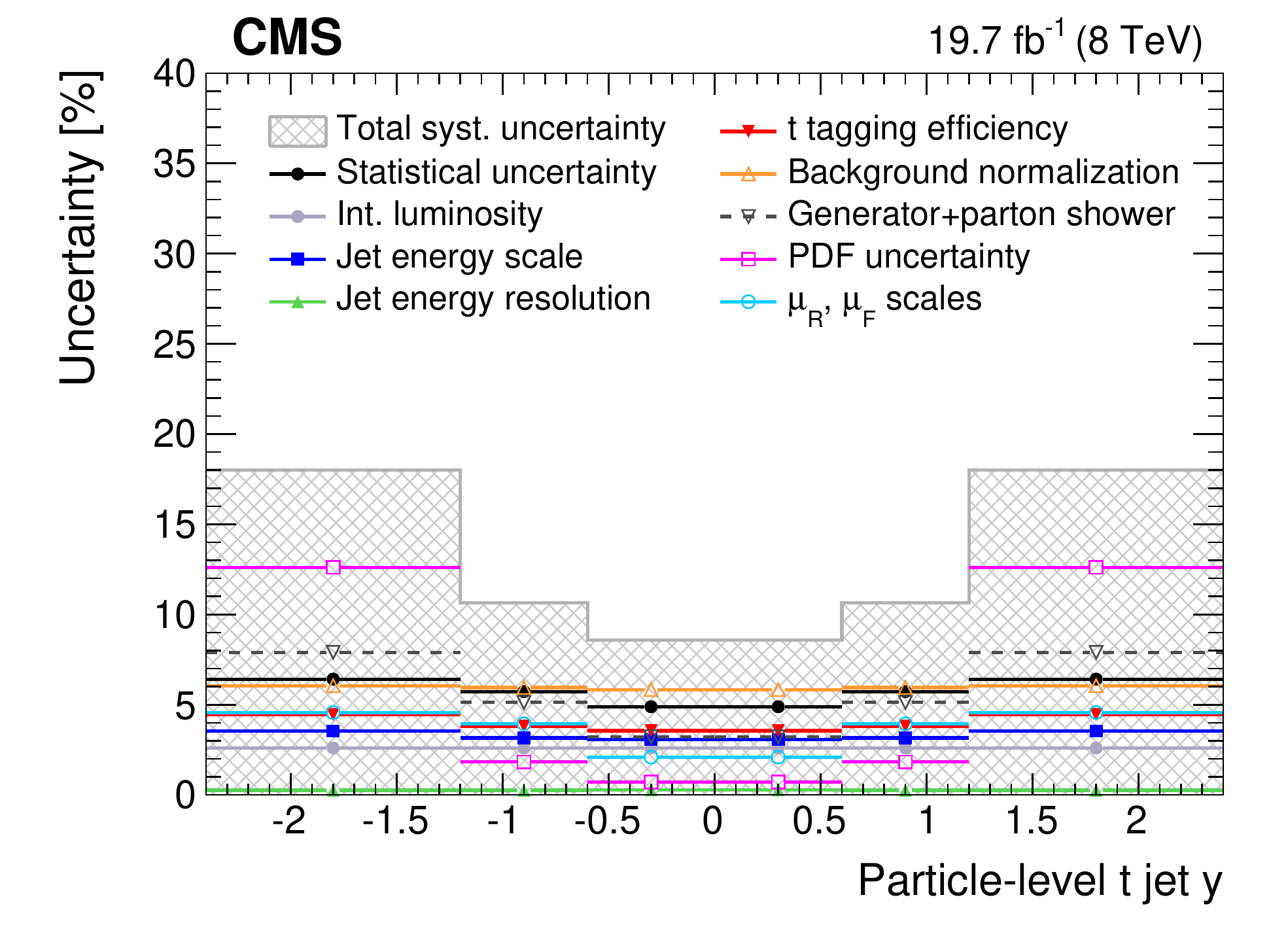}
\includegraphics[width=0.47\textwidth]{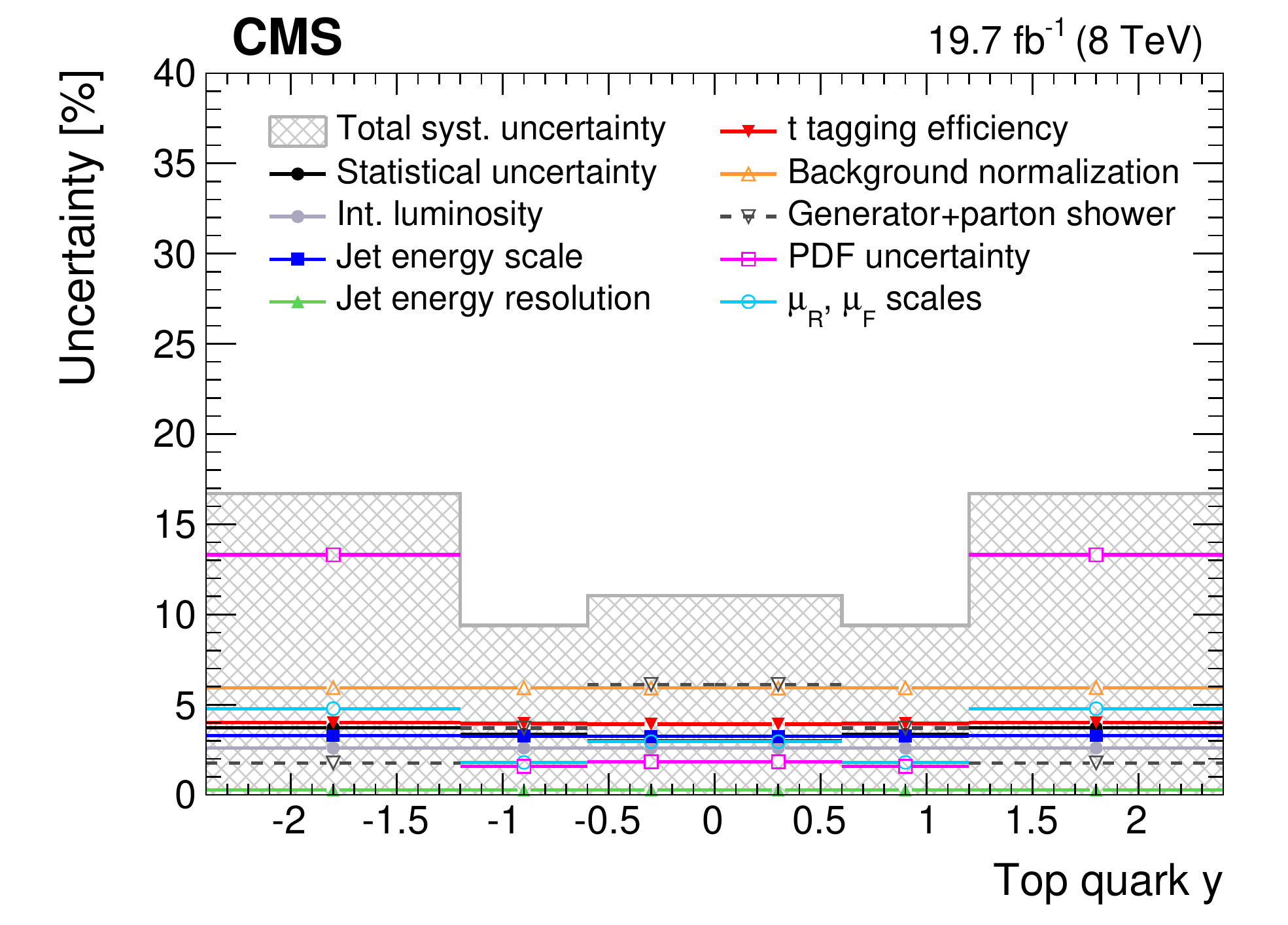}
\caption{\label{fig:unfold_relative_uncertainties_comb} Total systematic uncertainties (cross-hatched regions), as well as individual contributions and statistical-only uncertainties (points) in percent as a function of particle-level $\PQt$ jet $\pt$ (top left), parton-level top quark $\pt$ (top right), particle-level $\PQt$ jet $y$ (bottom left), and parton-level top quark $y$ (bottom right) for the differential cross section measurements. The horizontal bars on the points show the bin widths.}
\end{figure*}

The unfolded results at the particle and parton levels, including all experimental and theoretical uncertainties, are shown as a function of $\pt^{\PQt}$ and $y^{\PQt}$ as the data points in Fig.~\ref{fig:unfoldWithError_comb}, and the relative uncertainties are displayed in Fig.~\ref{fig:unfold_relative_uncertainties_comb}. As a consequence of bin migrations, the uncertainties at particle and parton level differ from the corresponding bin-by-bin uncertainties at detector level. 

The measured $\ttbar$ cross sections are listed in bins of $\pt^{\PQt}$ and $y^{\PQt}$ at the particle and parton levels in Table~\ref{tab:xsec_comb}. The measured cross sections are compared to the theoretical predictions from the {\PPsix}, {\MADPsix}, and {\MCATHsix} $\ttbar$ simulations, all normalized to the NNLO cross section~\cite{topNNLO}. Their values are also displayed in Fig.~\ref{fig:unfoldWithError_comb} and given in Table~\ref{tab:xsec_comb}. Also listed in Table~\ref{tab:xsec_comb} are the different relative uncertainties in the measurements, separated into the statistical uncertainty (Stat), the combined experimental uncertainty (Exp), the theoretical uncertainty (Th), and the total measurement uncertainty (Tot), all in percent.
The measured cross sections are lower than the predictions from {\PPsix} and {\MADPsix}, in particular for the high-$\pt^{\PQt}$ region, while {\MCATHsix} gives a better modeling of the data across the full $\pt^{\PQt}$ range. The differential cross sections are significantly overestimated for $\abs{y^{\PQt}} < 1.2$ by {\MADPsix} as compared to the data. The predictions of the $y^{\PQt}$ distributions by {\MCATHsix} and {\PPsix} agree with the data within the measurement uncertainties.

\begin{table*}[htb]
\topcaption{\label{tab:xsec_comb} Differential $\ttbar$ cross section in bins of $\pt$ and $y$ for the $\PQt$ jet at the particle level (top) and the top quark at parton level (bottom). The measurements are compared to predictions from the {\PPsix}, {\MADPsix}, and {\MCATHsix} simulations. The total relative uncertainty (Tot) in the measurements is separated into relative statistical (Stat), experimental (Exp), and theoretical (Th) components, all in percent.}
\centering
\cmsTable{\textwidth}{\begin{scotch}{ccccccccc}
\multirow{2}{*}{$\pt$ (\GeVns)} & \multicolumn{8}{c}{ $\rd\sigma/\rd\pt$ (fb/GeV) at particle level} \\[+1mm]
 & {Data} & {Stat (\%)} & {Exp (\%)} & {Th (\%)} & {Tot (\%)} & {\ \POWHEG} & {\MADGRAPH} & {\MCATNLO} \\
\hline
400--500 & 2.95 & 4.5 & 7.4 & 3.2 & 9.6 & 3.32 & 3.89 & 3.00 \\
500--600 & 1.29 & 4.5 & 8.4 & 8.6 & 13 & 1.52 & 1.77 & 1.25 \\
600--700 & 0.471 & 5.8 & 9.1 & 17 & 21 & 0.587 & 0.686 & 0.445 \\
700--800 & 0.166 & 7.9 & 11 & 16 & 22 & 0.222 & 0.249 & 0.185 \\
800--1200 & 0.029 & 9.7 & 15 & 37 & 41 & 0.038 & 0.039 & 0.025 \\
\hline
$y$ & \multicolumn{8}{c}{$\rd\sigma/\rd y$ (fb) at particle level} \\
\hline
$(-2.4, -1.2)$ & 27 & 6.4 & 8.3 & 16 & 19 & 34 & 33 & 27 \\
$(-1.2, -0.6)$ & 146  & 5.8 & 7.8 & 7.1 & 12 & 165  & 191 & 138 \\
$(-0.6, ~0.0)$ & 221  & 4.9 & 7.5 & 4.1 & 10 & 244  & 306 & 218 \\
$( 0.0,~0.6)$ & 221  & 4.9 & 7.5 & 4.1 & 10 & 252  & 303 & 215 \\
$( 0.6,~1.2)$ & 138  & 5.8 & 7.8 & 7.1 & 12 & 168  & 193 & 150 \\
$( 1.2,~2.4)$ & 26 & 6.4 & 8.3 & 16 & 19 & 35 & 33 & 28 \\
\end{scotch}
}
\null\vspace*{1em}
\cmsTable{\textwidth}{
\begin{scotch}{ccccccccc}
\multirow{2}{*}{$\pt$ (\GeVns)} & \multicolumn{8}{c}{$\rd\sigma/\rd\pt$ (fb/GeV) at parton level} \\[+1mm]
 & {Data} & {Stat (\%)} & {Exp (\%)} & {Th (\%)} & {Tot (\%)} & { \POWHEG} & {\MADGRAPH} & {\MCATNLO} \\
\hline
400--500 & 10.4 & 2.3 & 8.1 & 6.8 & 11 & 11.9 & 13.1 & 10.4 \\
500--600 & 2.74 & 2.3 & 9.0 & 10 & 14 & 3.25 & 3.64 & 2.63 \\
600--700 & 0.786 & 2.8 & 10 & 18 & 21 & 0.972 & 1.11 & 0.728 \\
700--800 & 0.254 & 3.7 & 12 & 16 & 20 & 0.324 & 0.363 & 0.256 \\
800--1200 & 0.036 & 4.5 & 13 & 30 & 33 & 0.049 & 0.050 & 0.033 \\
\hline
$y$ & \multicolumn{8}{c}{$\rd\sigma/\rd y$ (fb) at parton level} \\
\hline
$(-2.4, -1.2)$ & 83 & 3.7 & 7.9 & 14 & 17 & 94 & 88 & 82 \\
$(-1.2, -0.6)$ & 418  & 3.4 & 7.8 & 4.5 & 10 & 474  & 518  & 402 \\
$(-0.6, ~0.0)$ & 623  & 3.0 & 7.8 & 7.3 & 11 & 717  & 842  & 608 \\
$( 0.0, ~0.6)$ & 634  & 3.0 & 7.8 & 7.3 & 11 & 737  & 840  & 606 \\
$( 0.6, ~1.2)$ & 397  & 3.4 & 7.8 & 4.5 & 10 & 474  & 518  & 413 \\
$( 1.2, ~2.4)$ & 79 & 3.7 & 7.9 & 14 & 17 & 95 & 91 & 84 \\
\end{scotch}
}
\end{table*}

\begin{figure}[htbp]
\centering
\includegraphics[width=0.47\textwidth]{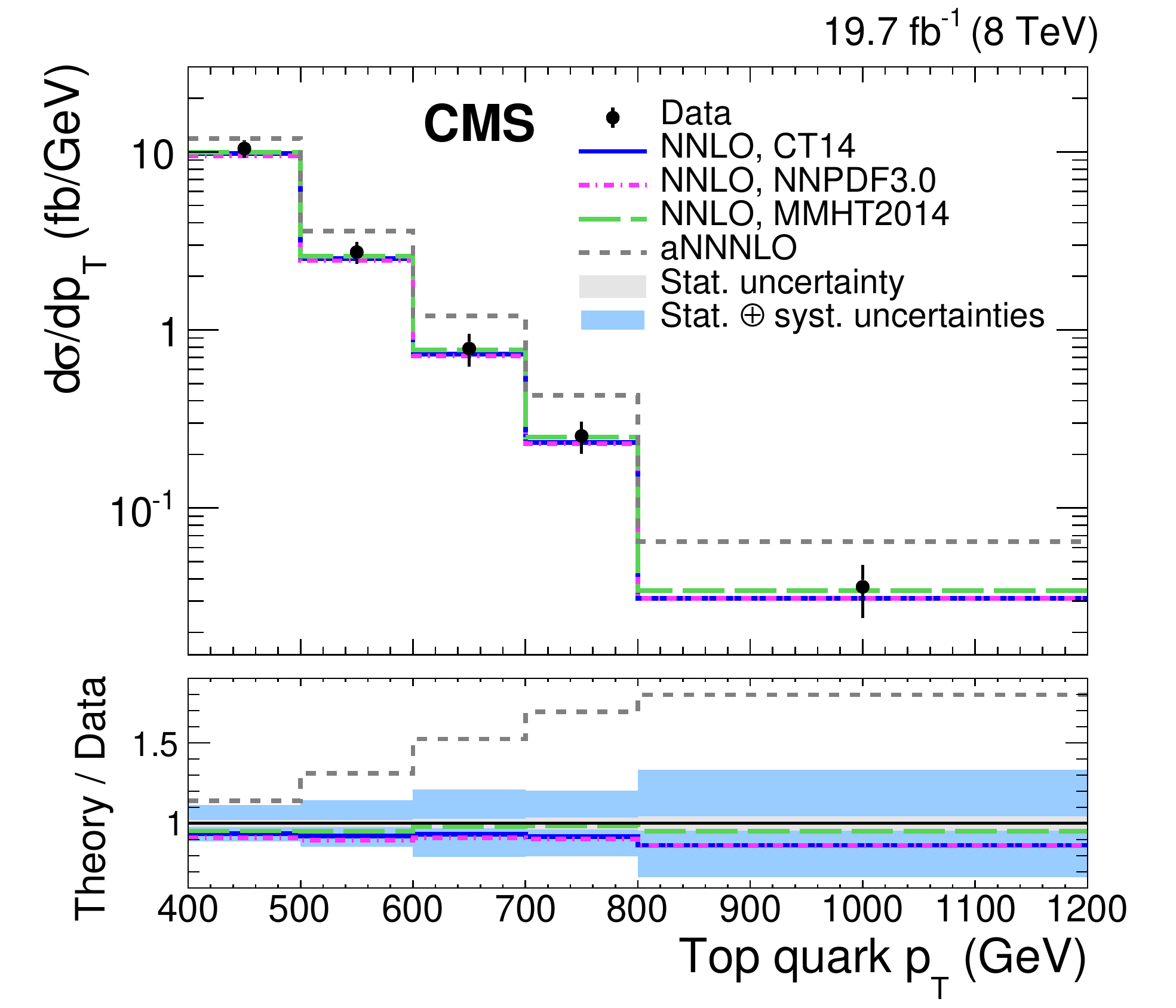}
\caption{\label{fig:unfoldWithError_nnlo} Differential $\ttbar$ cross section in bins of parton-level top quark $\pt$ including all systematic uncertainties. The measured cross section is compared to theoretical calculations at NNLO for three different PDF sets~\cite{Czakon:2016dgf} and at aNNNLO~\cite{Kidonakis:2014pja}. The lower plot shows the ratio of these theoretical predictions to the data. The statistical uncertainties are represented by the inner vertical bars with ticks and the light bands in the ratios. The combined uncertainties are shown as full vertical bars and the dark solid bands in the ratios.}
\end{figure}

The differential $\ttbar$ cross section measurement in bins of parton-level top quark $\pt$ is compared to different theoretical cross section calculations in Fig.~\ref{fig:unfoldWithError_nnlo}. Calculations of NNLO differential cross sections are extracted from Ref.~\cite{Czakon:2016dgf} for three different PDF sets (NNPDF3.0~\cite{Ball:2014uwa}, CT14~\cite{Dulat:2015mca}, and MMHT2014~\cite{Harland-Lang:2014zoa}). Approximate next-to-next-to-next-to-leading-order (aNNNLO) predictions corresponding to the results presented in Ref.~\cite{Kidonakis:2014pja} were provided by the author.
The NNLO calculations are in good agreement with the measurement across the full top quark $\pt$ range studied. Predictions for different PDF sets cannot be distinguished given the current measurement uncertainty but are all observed to be consistent with the data.
The aNNNLO calculation significantly overestimates the cross section, with an increasing disagreement with higher top quark $\pt$.
An additional check of the unfolding procedure is performed to confirm that the unfolding itself would support such a different $\pt$ spectrum. The {\PPsix} simulation is unfolded using response matrices derived from the same sample, but reweighting the distribution at detector level by a factor that corresponds to that required to match the aNNNLO prediction at parton level. The scaled and then unfolded simulation reproduces the aNNNLO prediction within the measurement uncertainty.

\section{Summary}
\label{sec:conclusion}

The first CMS measurement of the $\ttbar$ production cross section in the boosted regime has been presented. The integrated cross section, as well as differential cross sections as a function of the top quark $\pt$ and $y$, have been measured for $\pt^{\PQt} > 400\GeV$. The measurements use lepton+jets events, identified through an electron or a muon, a $\PQb$ jet candidate from the semileptonic top quark decay, and a $\PQt$ jet candidate from the top quark decaying to a hadronic final state. Backgrounds are modeled using simulations for the distributions, or a data sideband for multijet production. Background normalizations are extracted jointly with the signal yield and the $\PQt$~tagging efficiency using a maximum-likelihood fit.

{\sloppy
The integrated cross section measured for $\pt^{\PQt} > 400\GeV$ is $\sigma_{\ttbar} = 0.499 \pm 0.035\,(\text{stat+syst}) \pm 0.095 \thy \pm 0.013\lum\unit{pb}$ at particle level, and $\sigma_{\ttbar} = 1.44 \pm 0.10\,\text{(stat+syst)}  \pm 0.29\thy\pm 0.04\lum\unit{pb}$ at parton level, both corrected for the branching fraction of $\ttbar \to \Pe/\mu$+jets. The measurements are compared to the predicted cross section for this $\pt$ range from the {\PPsix} $\ttbar$ simulation assuming $\sigma_\text{tot} =  252.9$\unit{pb}, which provides a value of 0.580\unit{pb} at particle level and 1.67\unit{pb} at parton level. The cross section for this high-$\pt$ region is therefore found to be overestimated by 14\% in the {\PPsix} simulation, but is consistent within the uncertainties.\par}

Differential cross sections are also measured at both particle and parton levels. Background contributions are subtracted from the $\PQt$-tagged jet distributions to obtain the distribution for signal. This is unfolded first to the particle level to correct for signal efficiency, acceptance, and bin migrations to yield the cross section in bins of $\PQt$ jet $\pt$ and $y$ at particle level. The data are further unfolded to the parton level to extract the cross section in bins of top quark $\pt$ and $y$. The measurements are compared to predictions from different $\ttbar$ simulations. The {\PPsix} and {\MADPsix} simulations are observed to overestimate the cross section, in particular at high $\pt^{\PQt}$, while {\MCATHsix} results in a good modeling of the $\pt^{\PQt}$ spectrum. The {\PPsix} and {\MCATHsix} simulations model the $y^{\PQt}$ distributions well, while {\MADPsix} significantly overestimates the cross section for $\abs{y^{\PQt}} < 1.2$.
The results are compatible with those from the nonboosted CMS measurement~\cite{diff_xs_ttbar_resolved_8TeV} in the $\pt$ range where the two analyses overlap (400--500\GeV). The nonboosted measurement also observes an overestimate of the cross section for different MC generators in this $\pt$ range, most prominent for {\MADPsix}, and an improved modeling of the $\pt$ spectrum using {\HERWIG6} for the parton showering.
The measurement as a function of parton-level top quark $\pt$ is also compared to theoretical aNNNLO and NNLO calculations. While the aNNNLO prediction significantly overestimate the measurement, especially for high top quark $\pt$, the NNLO calculations are in good agreement across the full $\pt$ range studied.

The analysis presented in this paper extends the differential $\ttbar$ cross section measurement into the $\pt > 1\TeV$ range. These measurements will help improve the modeling of event generators in this high-$\pt$ range, an important regime for many new physics searches.

\section*{Acknowledgements}

\hyphenation{Bundes-ministerium Forschungs-gemeinschaft Forschungs-zentren} We congratulate our colleagues in the CERN accelerator departments for the excellent performance of the LHC and thank the technical and administrative staffs at CERN and at other CMS institutes for their contributions to the success of the CMS effort. In addition, we gratefully acknowledge the computing centers and personnel of the Worldwide LHC Computing Grid for delivering so effectively the computing infrastructure essential to our analyses. Finally, we acknowledge the enduring support for the construction and operation of the LHC and the CMS detector provided by the following funding agencies: the Austrian Federal Ministry of Science, Research and Economy and the Austrian Science Fund; the Belgian Fonds de la Recherche Scientifique, and Fonds voor Wetenschappelijk Onderzoek; the Brazilian Funding Agencies (CNPq, CAPES, FAPERJ, and FAPESP); the Bulgarian Ministry of Education and Science; CERN; the Chinese Academy of Sciences, Ministry of Science and Technology, and National Natural Science Foundation of China; the Colombian Funding Agency (COLCIENCIAS); the Croatian Ministry of Science, Education and Sport, and the Croatian Science Foundation; the Research Promotion Foundation, Cyprus; the Ministry of Education and Research, Estonian Research Council via IUT23-4 and IUT23-6 and European Regional Development Fund, Estonia; the Academy of Finland, Finnish Ministry of Education and Culture, and Helsinki Institute of Physics; the Institut National de Physique Nucl\'eaire et de Physique des Particules~/~CNRS, and Commissariat \`a l'\'Energie Atomique et aux \'Energies Alternatives~/~CEA, France; the Bundesministerium f\"ur Bildung und Forschung, Deutsche Forschungsgemeinschaft, and Helmholtz-Gemeinschaft Deutscher Forschungszentren, Germany; the General Secretariat for Research and Technology, Greece; the National Scientific Research Foundation, and National Innovation Office, Hungary; the Department of Atomic Energy and the Department of Science and Technology, India; the Institute for Studies in Theoretical Physics and Mathematics, Iran; the Science Foundation, Ireland; the Istituto Nazionale di Fisica Nucleare, Italy; the Ministry of Science, ICT and Future Planning, and National Research Foundation (NRF), Republic of Korea; the Lithuanian Academy of Sciences; the Ministry of Education, and University of Malaya (Malaysia); the Mexican Funding Agencies (BUAP, CINVESTAV, CONACYT, LNS, SEP, and UASLP-FAI); the Ministry of Business, Innovation and Employment, New Zealand; the Pakistan Atomic Energy Commission; the Ministry of Science and Higher Education and the National Science Center, Poland; the Funda\c{c}\~ao para a Ci\^encia e a Tecnologia, Portugal; JINR, Dubna; the Ministry of Education and Science of the Russian Federation, the Federal Agency of Atomic Energy of the Russian Federation, Russian Academy of Sciences, and the Russian Foundation for Basic Research; the Ministry of Education, Science and Technological Development of Serbia; the Secretar\'{\i}a de Estado de Investigaci\'on, Desarrollo e Innovaci\'on and Programa Consolider-Ingenio 2010, Spain; the Swiss Funding Agencies (ETH Board, ETH Zurich, PSI, SNF, UniZH, Canton Zurich, and SER); the Ministry of Science and Technology, Taipei; the Thailand Center of Excellence in Physics, the Institute for the Promotion of Teaching Science and Technology of Thailand, Special Task Force for Activating Research and the National Science and Technology Development Agency of Thailand; the Scientific and Technical Research Council of Turkey, and Turkish Atomic Energy Authority; the National Academy of Sciences of Ukraine, and State Fund for Fundamental Researches, Ukraine; the Science and Technology Facilities Council, UK; the US Department of Energy, and the US National Science Foundation.

Individuals have received support from the Marie-Curie program and the European Research Council and EPLANET (European Union); the Leventis Foundation; the A. P. Sloan Foundation; the Alexander von Humboldt Foundation; the Belgian Federal Science Policy Office; the Fonds pour la Formation \`a la Recherche dans l'Industrie et dans l'Agriculture (FRIA-Belgium); the Agentschap voor Innovatie door Wetenschap en Technologie (IWT-Belgium); the Ministry of Education, Youth and Sports (MEYS) of the Czech Republic; the Council of Science and Industrial Research, India; the HOMING PLUS program of the Foundation for Polish Science, cofinanced from European Union, Regional Development Fund; the Mobility Plus program of the Ministry of Science and Higher Education (Poland); the OPUS program of the National Science Center (Poland); MIUR project 20108T4XTM (Italy); the Thalis and Aristeia programs cofinanced by EU-ESF and the Greek NSRF; the National Priorities Research Program by Qatar National Research Fund; the Rachadapisek Sompot Fund for Postdoctoral Fellowship, Chulalongkorn University (Thailand); the Chulalongkorn Academic into Its 2nd Century Project Advancement Project (Thailand); and the Welch Foundation, contract C-1845.

\clearpage

\bibliography{auto_generated}

\cleardoublepage \appendix\section{The CMS Collaboration \label{app:collab}}\begin{sloppypar}\hyphenpenalty=5000\widowpenalty=500\clubpenalty=5000\input{TOP-14-012-authorlist.tex}\end{sloppypar}
\end{document}

%% file: TOP-14-012-authorlist.tex
\textbf{Yerevan Physics Institute,  Yerevan,  Armenia}\\*[0pt]
V.~Khachatryan, A.M.~Sirunyan, A.~Tumasyan
\vskip\cmsinstskip
\textbf{Institut f\"{u}r Hochenergiephysik der OeAW,  Wien,  Austria}\\*[0pt]
W.~Adam, E.~Asilar, T.~Bergauer, J.~Brandstetter, E.~Brondolin, M.~Dragicevic, J.~Er\"{o}, M.~Flechl, M.~Friedl, R.~Fr\"{u}hwirth\cmsAuthorMark{1}, V.M.~Ghete, C.~Hartl, N.~H\"{o}rmann, J.~Hrubec, M.~Jeitler\cmsAuthorMark{1}, A.~K\"{o}nig, M.~Krammer\cmsAuthorMark{1}, I.~Kr\"{a}tschmer, D.~Liko, T.~Matsushita, I.~Mikulec, D.~Rabady, N.~Rad, B.~Rahbaran, H.~Rohringer, J.~Schieck\cmsAuthorMark{1}, J.~Strauss, W.~Treberer-Treberspurg, W.~Waltenberger, C.-E.~Wulz\cmsAuthorMark{1}
% ...
\vskip\cmsinstskip
% ...
1:~~Also at Vienna University of Technology, Vienna, Austria\\
% ...
